\documentclass[twocolumn,showpacs,showkeys,preprintnumbers,amsmath,amssymb]{revtex4}

\usepackage{graphicx}  
\usepackage{dcolumn}   
\usepackage{bm}        

\usepackage{amssymb}
\usepackage{amsfonts}
\usepackage{latexsym}

\def\beq{\begin{eqnarray}}
\def\eeq{\end{eqnarray}}
\def\ln{\,\mbox{ln}\,}
\def\Det{\,\mbox{Det}\,}

\def\tr{\,\mbox{tr}\,}
\def\Tr{\,\mbox{Tr}\,}

\def\al{\alpha}
\def\be{\beta}
\def\ch{\chi}
\def\ga{\gamma}
\def\de{\delta}

\def\ep{\epsilon}

\def\la{\lambda}
\def\na{\nabla}

\def\rh{\rho}
\def\si{\sigma}

\def\ta{\tau}
\def\th{\theta}

\def\Ga{\Gamma}

\input{psfig}

\begin{document}

\preprint{hep-th/0307030}

\title{Conformal Quantum Gravity with the Gauss-Bonnet Term}

\author{Guilherme de Berredo-Peixoto}
\email{guilherme@fisica.ufjf.br}
\author{Ilya L. Shapiro}
\altaffiliation{On leave from Tomsk State
Pedagogical University, Russia.}
\email{shapiro@fisica.ufjf.br}
\affiliation{Departamento de F\'{\i}sica -- ICE, \\
Universidade Federal de Juiz de Fora, \\
Juiz de Fora, CEP: 36036-330, MG,  Brazil}


\begin{abstract}
The conformal gravity is one of the most important models of
quantum gravity with higher derivatives. We investigate the role
of the Gauss-Bonnet term in this theory. The coincidence limit of
the second coefficient of the Schwinger-DeWitt expansion is
evaluated in an arbitrary dimension $n$. In the limit $n=4$ the
Gauss-Bonnet term is topological and its contribution cancels.
This cancellation provides an efficient test for the correctness
of calculation and, simultaneously, clarifies the long-standing
general problem concerning the role of the topological term in
quantum gravity. For $n\neq 4$ the Gauss-Bonnet term becomes
dynamical in the classical theory and relevant at the quantum
level. In particular, the renormalization group equations in dimension
$n=4-\epsilon$ manifest new fixed points due to quantum effects of
this term.
\end{abstract}



\maketitle


\section{\label{1}Introduction}

At both classical and quantum levels, local conformal
symmetry plays a special role in theories of gravity and
their applications \cite{birdav,book,duff94}. One
of the most interesting issues is the violation of this
symmetry at the quantum level.
For the quantum theory of matter fields in curved space-time
the violation of conformal symmetry is related to the
well-known trace anomaly (see, e.g. \cite{birdav, duff94}
for the review). The important feature
of the conformal anomaly is its universality for the matter
(scalar, spinor and vector) fields which
contribute with the same sign to the two of three terms of
the anomaly. The opposite sign of the
contributions takes place for the unphysical higher derivative
scalars and fermions
\cite{rei,cofe}. In principle, one can choose the number
of these higher derivative fields in such a way that they
cancel the contributions of the matter fields. In this case
the conformal symmetry holds at the one loop level. The
cancellation of anomaly can not be achieved in the known
versions of conformal supergravity \cite{frts-sugra}, and
therefore the relation between the cancellation of conformal
anomaly and what is supposed to be the fundamental theory
(e.g. supergravity which may be a low-energy limit of the
(super)string/M - theory) remains unclear within the
semiclassical approach.

The violation of the conformal symmetry in quantum gravity
is much less studied. One of the simplest theories of gravity
which possesses local conformal symmetry is based on the
Weyl action $\,\,\int d^4x\sqrt{-g}C^2$, where
\beq
C^2 & = & C_{\mu\nu\al\be}C^{\mu\nu\al\be} \,=\,
R_{\mu\nu\al\be}R^{\mu\nu\al\be}
\,-\, 2 \,R_{\al\be}R^{\al\be} \nonumber  \\
& + & \frac13\,R^2\, \label{C2}
\eeq
is the square of the Weyl
tensor in $\,n=4\,$ dimensions. In order to provide
renormalizability, one has to include topological and surface
terms. In this way we arrive at the action
\beq
S_{W}\,=\,-\int
d^4x\sqrt{-g}\, \Big\{\,\,\frac{1}{2\la}\,C^2 \,+\,\eta\,E
\,+\,\tau\,\square R\,\Big\}\,. \label{Weyl}
\eeq
Here
\beq
E\,=\,R_{\mu\nu\al\be}R^{\mu\nu\al\be}
\,-\,4\,R_{\al\be}R^{\al\be}\,+\,R^2 \label{E}
\eeq
is the
integrand of the Gauss-Bonnet topological term. The action
(\ref{Weyl}) is conformal invariant, for it satisfies the
conformal Noether identity
\beq
-\,\,\frac{2}{\sqrt{-g}}\,
g_{\mu\nu}\, \frac{\de S_W}{\de g_{\mu\nu}}\,=\,0 \,.
\label{Noether}
\eeq

By dimensional reasons one can introduce into action
(\ref{Weyl}) an extra term $\,\,\th \cdot\int\sqrt{-g}\,R^2$.
However, this expression possesses only global and not local
conformal symmetry and hence it will not be included into the
action. In order to complete the story, let us notice that a
finite $\,\,\th \cdot\int\sqrt{-g}\,R^2\,$ term may be generated
as a quantum anomaly-induced correction, e.g. due to the
renormalization of the $\,\,\int\sqrt{-g}\square R\,\,$ term in
(\ref{Weyl}). The anomalous violation of local conformal symmetry
in the finite part of the one-loop effective action may produce
the non-conformal divergences beyond the one-loop level. This
effect has been investigated in \cite{hathrell} for the conformal
scalar field and there are no reasons to expect that the situation
for the conformal quantum gravity will be different.

The main purpose of the present paper is the one loop
renormalization
and renormalization group in the conformal quantum gravity.
The renormalization structure depends on the form of
divergences and corresponding counterterms.
According to the standard expectations,
despite the anomaly results from the one-loop renormalization,
one-loop divergences in conformal quantum gravity
must be conformally invariant. This
property holds in all known examples of conformal matter
fields, and one can expect that the same should be true for
the Weyl quantum gravity based on action (\ref{Weyl}).
The most natural result would be to meet the
renormalization of the coefficients $\,\eta,\,\la,\,\tau\,$
but not the $\,\,\int\sqrt{-g}\,R^2\,$-type counterterm.

From the first sight there is no much difference whether
the $\,\,\int\sqrt{-g}R^2\,$ term shows up already in the
one-loop divergences or only at higher loops. However, this
may be relevant for some applications of quantum gravity.
If non-conformal divergences
show up only at higher loops, the conformal symmetry
may be considered as a good approximation. For example,
the one-loop renormalizability of conformal gravity
provides the possibility of the successful realization
of the anomaly-induced inflation scheme (see the discussion
in \cite{anomayo}) in the presence of quantum gravity.
At the same time, if the non-conformal divergence emerges
at the one-loop order, the conformal symmetry can not be
considered a reasonable approximation, because the running
of the coefficient $\,\th\,$ will be much stronger in this
case. In this situation a quantized conformal matter on
curved classical background also can not be considered as an
approximation to the full theory involving quantum gravity
(see, e.g. discussion in \cite{birdav,wald}). Finally,
we need to be sure whether the non-conformal divergence
is present at the one loop level in the Weyl quantum
gravity.

The first explicit derivation of the one-loop divergences
in Weyl quantum gravity has been performed by Fradkin
and Tseytlin \cite{frts82} in the framework of background
field method, properly modified for the higher derivative
theories \footnote{Let us notice that the main purpose
of \cite{frts82} was the one-loop calculation in the more
general theory including also $\,\,\int\sqrt{-g}R^2$,
Einstein-Hilbert and cosmological terms. Such calculation
has been first performed by Julve and Tonin in \cite{julve}
(see also \cite{avba} for the consequent calculation).}.
The $\,\,\int\sqrt{-g}R^2$-type
divergence has been encountered and the conformal invariance
of the counterterms has been achieved through the use of
the special procedure of conformal regularization. This
regularization is nothing but the specific reparametrization
of background metric invented earlier in \cite{truffin}
(see also \cite{frvi} and \cite{bush86}).
According to this procedure
metric $\,g_{\mu\nu}\,$ has to be replaced by conformal
metric $\,\tilde{g}_{\mu\nu}=g_{\mu\nu}\,P^2[g_{\mu\nu}]$,
where scalar metric-dependent quantity $\,P[g_{\mu\nu}]\,$
is defined as a solution of the equation
\beq
\square P = \frac16\,R\,P\,.
\label{P}
\eeq

When performing a local conformal transformation of
the original metric
\beq
&&g_{\mu\nu}\,\to\,g^\prime_{\mu\nu}\,=\,g_{\mu\nu}\,e^{2\si(x)}
\,,\quad \mbox{the quantity} \nonumber
\\ && P[g_{\mu\nu}] \quad \mbox{transforms as} \quad
P\,\to\,P^\prime = P \,e^{-\si(x)}\,, \nonumber
\eeq
such that the metric $\,\tilde{g}_{\mu\nu}\,$ remains
invariant. Another important property of the metric
$\,\tilde{g}_{\mu\nu}\,$ is that the corresponding scalar
curvature is zero $\,\tilde{R}=R(\tilde{g}_{\mu\nu})=0$.
Therefore, after the original metric $\,g_{\mu\nu}\,$
is replaced by $\,\tilde{g}_{\mu\nu}$, the divergent
$\,\,\int\sqrt{-g}R^2\,\,$ counterterm disappears
and the expected invariant form of divergences gets
restored. The procedure of conformal \lq\lq regularization"
has been generalized for the conformal quantum gravity
coupled to conformal quantum matter fields in \cite{bush86}.

Is it correct to consider the replacement
$\,g_{\mu\nu}\,\to\,\tilde{g}_{\mu\nu}\,$ as a kind of
conformal regularization for the divergent part of
the effective action of quantum gravity?
It is easy to see that this procedure eliminates
also the anomaly in the finite part of the effective
action \footnote{In fact, the conformal reparametrization
has been originally designed for this purpose
\cite{truffin}.}.
Therefore, this choice of background
metric does not fit with numerous applications of
conformal anomaly which we know. Furthermore, despite
the choice of $\,\tilde{g}_{\mu\nu}\,$ as a
background metric is mathematically consistent, it is
not very appealing because, in particular, it eliminates
the Einstein-Hilbert action. Consequently, the theory
based on this metric may not have a consistent
non-relativistic limit.
In general, the whole procedure
looks as an artificial addition to the background field
method. If we really want to learn the role of the conformal
symmetry in quantum gravity, it is important to know
whether the appearance of the $\,\,\int\sqrt{-g}R^2\,\,$
counterterm is a calculational error or it
is caused by inconsistency of the background
field method applied to Weyl quantum gravity. The last
option has been partially explored in \cite{shja},
where the possible conflict between diffeomorphism
and conformal gauge fixing conditions has been discussed.
It turned out that counterterm $\,\int\sqrt{-g}R^2\,$
is gauge fixing independent, exactly as the renormalization
of the terms $\,\int\sqrt{-g}C^2\,$ and $\,\int\sqrt{-g}E$.
On the opposite, counterterm $\,\int\sqrt{-g}\square R\,$
is not protected from the gauge fixing dependence and can
be modified or even eliminated by the appropriate choice
of the parameters of gauge fixing.

The second explicit derivation of the one-loop divergences
in Weyl quantum gravity has been performed by Antoniadis,
Mazur and Mottola \cite{amm} using methods developed
in \cite{frts82}. The correctness of the $\be$-functions for
the coefficients $\,\la\,$ and $\,\eta\,$ calculated in
\cite{frts82} has been confirmed. At the same time,
the paper in Ref. \cite{amm}
did not meet the suspicious $\,\,\int\sqrt{-g}R^2\,$
counterterm. Indeed, this result coincides with our
general expectations discussed above, but the situation
with the two conflicting results does not look acceptable.
In what follows we shall perform a more general quantum
calculation using dimensional regularization and starting
from the action (\ref{Weyl}). In this way we will be able,
in particular, to check the previous calculations \cite{frts82}
and \cite{amm}.
Even more important may be that we shall achieve better
understanding of the role of the
Gauss-Bonnet term in quantum gravity in $n=4$ and
$n=4-\ep$ dimensions.

As far as one of our objectives is to perform a very
complicated calculation in conformal quantum gravity,
we have to provide maximal safety with respect to
possible calculational errors. For this end we shall
use a new way of organizing calculations, which
guarantees an efficient automatic verification of our
result.
Simultaneously, we shall resolve another old-standing
problem of quantum gravity. In the well-known paper
\cite{capkim},
Capper and Kimber noticed that the Gauss-Bonnet term
may, in principle, play a significant role in quantum
gravity. Usually
this term is disregarded because it is topological and
does not affect the classical equations of motion. However,
this conclusion is true only if the Bianchi identity is
satisfied. This implies the diffeomorphism invariance
of the theory. However, when the theory is quantized
through the Faddeev-Popov procedure, the diffeomorphism
invariance is broken and the vector space extends
beyond the physical degrees of freedom. In other words,
after quantization not only the spin-2,
but also the spin-1 and spin-0 components
of the quantum metric become relevant, and the topological
term may produce new vertices of interaction between
these components. As a result, the quantum-gravitational
loops may be, in principle, affected by the
presence of the topological term. Of course, this output
does not look probable, because if we include the
topological term into the classical action, the
gauge-fixing condition should modify and eventually
compensate the new vertices. But this is a sort of
believes which are always good to verify. Such verification
is one of the purposes of the present paper.

We shall perform the one-loop calculation
starting from the full action (\ref{Weyl}), taking
the topological term into account. As it was predicted in
\cite{capkim}, the contributions of this term penetrate
all vertices or, in other words, all elements of the
background field method technique. However, despite many
intermediate formulas strongly depend on the coefficient
$\,\eta\,$, this dependence completely disappears in the
final expression for the divergent part of the effective
action in the $n\to 4$ limit. This cancellation means
negative answer to the hypothesis raised by Capper and
Kimber \cite{capkim}.
Moreover, it provides
a very strong test for the correctness of the calculation.
On the other hand, the $\eta$-dependence is present
in the $\,n\neq 4\,$ expression. Therefore,
derivation of the relevant part of the effective action
in an arbitrary dimension $\,n\neq 4$
opens the way for constructing the complete $\,4-\ep\,$
renormalization group equations in the conformal quantum
gravity theory (\ref{Weyl}). As it will be shown below,
the quantum effect of the Gauss-Bonnet term leads to
new fixed points, which have no analogs in the
$\,n=4\,$ case.

The paper is organized as follows. In the next section
we shall briefly describe the Lagrange quantization of
theory (\ref{Weyl}). One can find a detailed description
of this subject in Refs. \cite{stelle,frts82,Christen,book}.
In section \ref{3} the details of the bilinear expansion
of higher-derivative gravity are presented.
Some of the bulky expressions corresponding to this
section are collected in Appendix A.
Our expansions are more general than the ones
which were known before \cite{julve,Christen,book},
because they
are performed for all higher derivative terms including
$\,\sqrt{-g}E\,$
and without taking into account the conformal gauge
fixing condition. This enables one, in principle, to
derive divergences not only for the conformal case,
but also for the general higher derivative quantum gravity
\cite{julve,frts82,avba}. In the present paper we perform
only the calculation for the Weyl theory and expect to
report the results for the general case later on.
In section \ref{4} we derive the coincidence limit of the
$\,a_2(x,x^\prime)\,$ coefficient of the Schwinger-DeWitt
expansion. The expression is obtained for the general
$\,n$-dimensional space-time, in order to see the effect
of the Gauss-Bonnet term more explicitly. After that, we
derive the divergences of the Weyl gravity at $\,n\to 4$
and establish their independence on the
parameter $\,\eta$. In section \ref{5} the renormalization
group in the $\,4-\ep\,$ dimensions is considered, and
a number of new UV-stable and UV-unstable fixed points
(due to the quantum effects of the topological term) are
described. In the
course of the calculations in sections \ref{4},\ref{5}
we use the
computer algebra program MAPLE (see, e.g. \cite{MAPLE}).
Finally, in the last section we draw our conclusions and
discuss the possible form of the non-conformal finite
contributions to the one-loop effective action.

\section{\label{2}Quantization and gauge-fixing dependence}

The quantum gravity
calculation in the background field method (see, e.g.
\cite{book} for the introduction) implies the special
parametrization of the metric
\beq
g_{\mu\nu} \,\to\, g^\prime_{\mu\nu}
\,=\,g_{\mu\nu}+h_{\mu\nu}\,.
\label{background}
\eeq
In the {\it r.h.s.} of the last formula $\,g_{\mu\nu}\,$ is
the background metric and $\,h_{\mu\nu}\,$ is the
quantum field (integration variable in the path integral).
The 1-loop contribution $\,\bar{\Ga}^{(1)}\,$ to the
effective action of quantum gravity is defined as follows
\cite{frts82}
\beq
\bar{\Ga}^{(1)}[g_{\mu\nu}]
& = & \frac{i}{2}\,\ln\,\Det \hat{\cal{H}}
\,\, -\,\,\frac{i}{2}\,\ln\,\Det \,Y^{\al\be} - \nonumber \\
& - & i\,\ln\,\Det \hat{\cal{H}}_{gh}\,,
\label{e}
\eeq
where $\,\hat{\cal{H}}\,$ is the bilinear (in quantum fields)
form of the action (\ref{Weyl}) together with the gauge
fixing term
\beq
S_{GF}\,=\,\mu^{n-4\,}\int
d^nx\sqrt{-g}\,\,\chi_\al\,Y^{\al\be}\,\chi_\be\,.
\label{gauge}
\eeq
The operator $\,\hat{\cal{H}}_{gh}\,$ is a bilinear
form of the
action of the Faddeev-Popov ghosts and $\,\mu\,$ is the
dimensional constant (renormalization parameter in the
dimensional regularization). The expression (\ref{e})
includes also $\,\ln \Det Y^{\al\be}\,$, where
$\,\,Y^{\al\be}\,\,$ is the weight function. In
the case of the higher derivative gravity theory this term
gives relevant contribution to the effective action, because
$\,Y^{\al\be}\,$ is a second order differential operator
\cite{frts82}.

Introducing the gauge fixing term (\ref{gauge}) one is
fixing the diffeomorphism invariance. However, in the theory
under consideration this is not sufficient, because there
is another classical symmetry - local conformal invariance,
which leads to a degeneracy
even after the term (\ref{gauge}) is introduced. Hence one
has to choose the second gauge fixing condition.
Following Fradkin and Tseytlin \cite{frts82}, we fix the
conformal symmetry by
imposing the constraint $\,h=h_\mu^\mu=0$. The interference
between the two gauge fixing conditions may take place because
the term (\ref{gauge}) breaks the conformal symmetry in the
background fields sector \cite{shja}. However, this breaking
can not lead to the non-conformal counterterms, because
the last can be shown insensitive to the choice of the
gauge fixing condition. The general gauge fixing condition
(here we restrict our attention to the linear background
gauges) has the form
\beq
\ch^\mu & = & \na_{\la}h^{\la\mu}+\beta\na^\mu h\ \nonumber \\
Y_{\mu\nu} & = & \frac{1}{\al}\,\Big( g_{\mu\nu}\Box +
\ga\na_\mu\na_\nu -\de\na_\nu\na_\mu + \nonumber \\
& + & p_1R_{\mu\nu}
+p_2Rg_{\mu\nu}\Big)\,,
\label{fixing}
\eeq
where $\,\al,\,\be,\,\ga,\,\de,\,p_1,\,p_2\,$ are arbitrary
parameters. The action of the Faddeev-Popov ghosts has the
form
\beq
S_{gh} \,=\,
\int d^4 x\sqrt{-g}\,\,{\bar C}^\mu\,
\left({\cal H}_{gh}\right)_\mu^\nu\, C_\nu\,,
\label{ghost action}
\eeq
where
\beq
\hat{\cal H}_{gh}=\left({\cal H}_{gh}\right)_\mu^\nu
= - \de_\mu^\nu\,\Box - \na^\nu \na_\mu
- 2\be \na_\mu \na^\nu\,.
\label{ghost operator}
\eeq
The parameter  $\,\be\,$ is fixed
in the conformal case due to the conformal symmetry
condition $\,h_\mu^\mu=0$, hence $\,\be=-1/n$.
Other parameters may take different values and their
choice may influence, in principle, the one-loop
divergences. The general analysis \cite{shja} shows that
the $\,C^2,\,E\,$ and $\,R^2\,$ counterterms can not depend
on these parameters while the $\,\square R\,$-type
counterterm may have such dependence. In what follows
we shall use these
data extensively, namely we will not pay attention
to the irrelevant $\,\,\int\sqrt{-g}\,\square R\,\,$
counterterm and,
on the other hand, we shall choose the gauge fixing parameters
$\,\al,\,\ga,\,\de,\,p_1,\,p_2\,$ such that the calculation
of other counterterms becomes simpler. Let us remark that
the dependence on the parameters $\,p_1,\,p_2\,$ has been
explored and found irrelevant in \cite{avba} for the
non-conformal version
of the higher derivative quantum gravity.

\section{\label{3}Bilinear expansion}

\qquad
The action (\ref{Weyl}) includes only
higher derivative conformal invariant and surface terms.
There are no
Einstein-Hilbert, cosmological and $\,\int\sqrt{-g}\,R^2\,$
terms in the action, because none of them possesses local
conformal symmetry.
But for the sake of completeness, the
bilinear expansions for all these terms will be given too.
The parametrization of the quantum metric $\,h_{\mu\nu}\,$
will be chosen according to (\ref{background}).
Let us remark that the relevant divergences in the theory
(\ref{Weyl})
are independent of the choice of parametrization for the
quantum metric \cite{shja}. When making the expansions of the
elements of the gravitational action, we keep in mind that
the relevant terms are of the second order in  $\,h_{\mu\nu}$.
Hence we shall pay the main attention to this order of the
expansion. In what follows we indicate all quantities
constructed from the total metric $\,g^\prime_{\mu\nu}\,$
by prime (e.g. $\,{g^\prime}^{\mu\nu},$ $\,\sqrt{-g^\prime},$
$\,{\Ga^\prime}^\ga_{\mu\nu}$,
$\,{R^\prime}^\al_{\,\,\,\mu\be\nu}\,$
etc), and reserve simpler notations (e.g. $\,g^{\mu\nu}$,
$\,\sqrt{-g},$ $\,\Ga^\ga_{\mu\nu}$,
$\,R^\al_{\,\,\,\mu\be\nu}\,$
etc) for the quantities constructed from the background
metric $\,g_{\mu\nu}$.

For $\,{g^\prime}^{\mu\nu}\,$ and
$\,\sqrt{-g^\prime}\,$ the expansions can be presented as
\beq
{g^\prime}^{\mu\nu} & = & g^{\mu\nu}\,-\,h^{\mu\nu}
\,+\,h^\mu_\la\,h^{\la\nu}
\,-\,h^\mu_\la\,h^\la_\tau\,h^{\tau\nu}\,+\,...\, \nonumber \\
& = & g_{(0)}^{\mu\nu}\,+\,g_{(1)}^{\mu\nu}
\,+\,g_{(2)}^{\mu\nu}\,+\,g_{(3)}^{\mu\nu}\,+\,...\,
\label{expansion 2} \eeq and \beq \sqrt{-{g^\prime}} =
\sqrt{-g}\,\Big(\,1 + \frac12\,h +\frac18\,h^2
-\frac14\,h_{\mu\nu}h^{\mu\nu}+ ...\Big).
\label{expansion 3}
\eeq
For the coefficients of the affine connection, using
(\ref{expansion 2}), we arrive at the expansion
\beq
{\Ga^\prime}^\al_{\mu\nu}=\Ga^\al_{\mu\nu}+
\de\Ga^\al_{\mu\nu}\,,\,{\rm where}\, \de\Ga^\al_{\mu\nu}
=\sum_{n=1}^{\infty}\de\Ga^{(n)}\,^\al_{\mu\nu}.
\label{expansio 4}
\eeq
Here the tensors $\,\de\Ga^{(n)}\,^\al_{\mu\nu}\,$ are
given by the expressions
\beq
\de\Ga^{(n)}\,^\al_{\mu\nu}=\frac12\,g_{(n-1)}^{\al\be}
\Big(\na_\mu h_{\be\nu}+\na_\nu h_{\be\mu} -\na_\be
h_{\mu\nu}\Big)\,.
\label{expansion 55}
\eeq
For the curvature
tensor one can establish the following general expression: \beq
{R^\prime}^\al_{\,\,\be\mu\nu} & = & R^\al_{\,\,\be\mu\nu}
\,+\,\na_\nu\,\de\Ga^\al_{\be\mu}
\,-\,\na_\mu\,\de\Ga^\al_{\be\nu}
\,+\,\de\Ga^\la_{\be\nu}\,\de\Ga^\al_{\la\mu}- \nonumber \\
& - & \de\Ga^\la_{\be\mu}\,\de\Ga^\al_{\la\nu}\,=
R^\al_{\,\,\be\mu\nu}
+ \sum_{n=1}^\infty R^{(n)}\,^\al_{\,\,\be\mu\nu}\,.
\label{expansion 5}
\eeq
In the first and second orders in the quantum metric,
$\,h_{\mu\nu}\,$, we obtain the following expressions
for the Riemann tensor:
\beq
R^{(1)}\,^\al_{\,\,\be\mu\nu}&=&
\frac12\,\Big(\na_\mu\na_\be h^\al_\nu -
\na_\nu\na_\be h^\al_\mu+ \na_\nu\na^\al h_{\be\mu} -
\nonumber
\\
& - & \na_\mu\na^\al h_{\be\nu}\,+
R^\al_{\,\,\la\mu\nu}\,h^\la_\be
-R^\la_{\,\,\be\mu\nu}\,h^\al_\la\Big)\,,
\nonumber
\\
\nonumber
\\
R^{(2)}\,^\al_{\,\,\be\mu\nu} & = &
\frac12\,h^{\al\la}\,\Big(\na_\mu\na_\la h_{\nu\be}
+\na_\nu\na_\be h_{\mu\la}+ \nonumber
\\
& + & \na_\nu\na_\mu h_{\la\be}\Big)+
\frac14\,\Big[
\na_\mu h^{\al\la}\,(\na_\la h^{\nu\be} -
\nonumber
\\
& - & \na_\be h^{\nu\la}-\na_\nu h^{\be\la}) +
\na_\be h^\la_\nu\,(\na_\la h^\al_\mu -
\nonumber
\\
& - & \na^\al h_{\la\mu})+
\na_\nu h^\la_\be\,(\na_\la h^\al_\mu -\na^\al h_{\la\mu}) +
\nonumber
\\
& + & \na^\la h_{\mu\be}\,(\na_\la h^\al_\nu -\na^\al h_{\la\nu})
\Big] \,- (\mu \leftrightarrow \nu)\,. \nonumber \\
\label{expansion 7}
\eeq
For the Ricci tensor, similar expansions have the form
\begin{widetext}
\beq
R^{(1)}_{\mu\nu} & = &
\frac12\,\left(\na_\la\na_\mu h^\la_\nu
+ \na_\la\na_\nu h^\la_\mu
- \na_\mu\na_\nu h - \Box h_{\mu\nu}\right)\,,
\nonumber
\\
R^{(2)}_{\mu\nu} & = &
\frac12\,h^{\al\be}\,
\Big(\na_\al\na_\be h_{\mu\nu}
+ \na_\mu\na_\nu h_{\al\be} - \na_\al\na_\mu h_{\nu\be}
- \na_\al\na_\nu h_{\mu\be}\Big)\,+
\frac12\,\na_\al h^{\al\be}\,
\Big(\na_\be h_{\mu\nu}- \na_\mu h_{\nu\be}-
\na_\nu h_{\mu\be}\Big) + \nonumber \\
& + & \frac12\,\na_\al h_{\mu\be}\,\Big(\na^\al h_{\nu}^\be
\,-\,\na^\be h_{\nu}^{\al}\Big)\,+
\frac14\,\na_\mu h^{\al\be}\,\na_\nu h_{\al\be}+
\frac14\,\na^\be h\,
\Big(\na_\mu h_{\nu\be} + \na_\nu h_{\mu\be}
-\na_\be h_{\mu\nu}\Big)\, .
\label{expansion 9}
\eeq
\end{widetext}
For the scalar curvature we meet the following expansions:
$$
R^{(1)} =
\na_\mu\na_\nu h^{\mu\nu}\,-\,\Box h\,-\,
h_{\mu\nu}\,R^{\mu\nu}\,,
$$
\beq
R^{(2)}&=& h^{\al\be}(\na_\al\na_\be h + \Box h_{\al\be}
- \na_\al\na_\la h^\la_\be - \na_\la\na_\al h^\la_\be)+ \nonumber \\
& + & \na_\al h^{\al\la}\,(\na_\la h - \na_\be h^\be_\la) -
\frac14\,\na_\la h\,\na^\la h +
\nonumber \\
& + & h_{\al\la}\,h_\be^\la\,R^{\al\be}+
\frac34\,\na_\la h_{\al\be}\,\na^\la h^{\al\be}
- \nonumber \\
& - & \frac12\,\na_\al h_{\la\be}\,\na^\be h^{\al\la}\,.
\label{expansion 11}
\eeq

With these expansions in hands, we can derive the part of
the action quadratic in the quantum fields. It proves useful
to consider an alternative version of the action (\ref{Weyl})
\beq
S_{W}(n) & = & -\mu^{(n-4)}\int d^nx\sqrt{-g}\,
\left\{\, x\, R_{\mu\nu\al\be}^2 \,+\,y\, R_{\mu\nu}^2
\,+\, \right.\nonumber \\
& + & \left. z\,R^2\,+\,\tau\square R\,\right\}\, ,
\label{higher}
\eeq
where the new parameters $x$, $y$ and $z$ are related
to $\,\eta\,$ and $\,\la\,$ as follows:
$$
x = \frac{1}{2\,\lambda}\,+\,\eta\,,\qquad
y = -\,{\frac {2}{\left (n-2\right )\lambda}}-4\,\eta\, ,
$$
\beq
z = \eta + {\frac {1}{\la\,(n-1)(n-2)}} \,.
\eeq

After some algebra we arrive at the formula
\beq
S^{(2)} & = & -\mu^{(n-4)}\int d^n x\,\Big\{
      x\Big(\sqrt{-g}\,R^2_{\mu\nu\al\be}\Big)^{(2)}+
\nonumber \\
& + & y\Big(\sqrt{-g}\,R^2_{\mu\nu}\Big)^{(2)} \,+\,
z\Big(\sqrt{-g}\,R^2\Big)^{(2)}\Big\} \,,
\label{expansion}
\eeq
where the complicated expressions for the bilinear
forms are collected in the Appendix A.

Starting from the expression (\ref{expansion}), and using
(\ref{Riemann}),(\ref{Ricci}) and (\ref{scalar}), one can
easily find the bilinear form of the action (\ref{higher}).
The operator
$\,\hat{\cal{H}}\,$ depends on the gauge fixing term
(\ref{gauge}). The gauge fixing parameters
$\,\al,\,\be,\,\ga,\,\de,\,p_1,\,p_2\,$ in (\ref{fixing})
will be
chosen in such a way that the operator takes the most simple,
minimal form \beq \hat{\cal{H}}\,=\,{\hat K}\,\square^2+{\cal
O}(\na^2)\,, \label{minimal high} \eeq where $\,{\hat K}\,$
is a non-degenerate $c$-number operator. The two  of the
possible non-minimal fourth derivative terms
$\,g_{\mu\nu}\na_\al\Box\na_\be\,$ and
$\,g_{\al\be}\na_\mu\Box\na_\nu\,$ vanish due to the
conformal gauge fixing condition $\,h_\mu^\mu=0$. The
simplest choice of the parameters providing the cancellation
of the remaining non-minimal fourth-derivative structures
$\,\na_\mu\na_\nu\na_\al\na_\be\,$
and \ $\,g_{\nu\be}\na_\mu\Box\na_\al\,$ is the following:
\beq
&&\al = \frac{2}{y+4x}\,, \qquad
\ga = \frac{2x-2z}{y+4x}\,, \nonumber \\
&&\de=1\,,\qquad p_1=p_2=0\,,
\label{minimal}
\eeq
with $\be=-1/n$
defined by the conformal gauge fixing (as we already
noticed). Let us remark that this
\lq\lq mi\-ni\-mal" choice of the gauge fixing is sensitive to the
introduction of the Gauss-Bonnet term, as we expected. On the
other hand, if we fix the value of $\,\eta\,$ such that the sum of
the Weyl term and the topological term gives
\beq
C^2\,-\,E\,=\,2W\,=\,2\Big(R_{\mu\nu}^2-\frac13\,R^2\,\Big),
\label{W}
\eeq
the gauge fixing condition (\ref{minimal}) coincides with
the one of \cite{frts82,amm}.

After some algebraic calculations and using the commutators
(\ref{commutator 2}), we find
$$
[S+S_{gf}]^{(2)} = h^{\mu\nu}\,\,\hat{\cal H}\,\,h^{\al\be}\,,
\qquad \mbox{where}
$$
\beq
\hat{\cal H}
\,=\,\hat{K}\Box ^2+\hat{D}^{\rh\la}\na_\rh\na_\la +
\hat{N}^\mu\na_\mu -(\na_\mu \hat{Z}^\mu)+\hat{W}
\nonumber \\
\label{bilinear}
\eeq
and $\,\hat{K}$, $\,\hat{D}^{\rh\la}$, $\,\hat{N}^\mu\na_\mu$,
$\,\na_\mu \hat{Z}^\mu\,$ and $\,\hat{W}\,$ are matrices
in the $\,h^{\mu\nu}$-space. In order to derive an
explicit form of $\,\hat{N}^\mu\,$ and $\,\hat{Z}^\mu\,$
one has to extend the derivation of bilinear expressions
of the Appendix A.
However, the derivation of these quantities does not have
much sense, because the terms $\,\hat{N}^\mu\na_\mu\,$
and $\,(\na_\mu \hat{Z}^\mu)$ may be safely disregarded.
The reason is that both expressions $\,\hat{N}^\mu\,$ and
$\,\hat{Z}^\mu\,$ are covariant derivatives of
curvatures. Therefore they may contribute only to the
irrelevant gauge-fixing dependent
$\,\int\sqrt{-g}\square R$-type counterterm, which we are
not calculating here. Below we shall simply set both terms
to zero.

Let us introduce the useful notation
$$
\bar{\de}_{\mu\nu ,\al\be}
\,=\,\de_{\mu\nu ,\al\be} \,-\,\frac{1}{n}
\,g_{\mu\nu}\,g_{\al\be}
$$
for the projection operator into the traceless sector of the
$\,h^{\mu\nu}\,$ space. Here, as usual,
$$
\de_{\mu\nu ,\al\be}\,=\,\frac12\,\big(g_{\mu\al}\,g_{\nu\be}
+ g_{\mu\be}\,g_{\nu\al}\big)\,.
$$
Since we assume the conformal gauge fixing condition $\,h=0$,
the tensor $\,\bar{\de}_{\mu\nu\, ,\,\al\be}\,$ plays the
role of the identity matrix. Without this condition
the identity matrix is $\,\de_{\mu\nu\,,\,\al\be}\,$.
Taking the conformal gauge into account, we find
\beq
(\hat{K})_{\mu\nu\, ,\,\al\be}\, & = & \,
\Big(\frac{y}{4}+x\Big)\,\bar{\de}_{\mu\nu\, ,\,\al\be}\,,
\label{K}
\eeq
\begin{widetext}
\beq
(\hat{D}^{\rh\la})_{\mu\nu ,\al\be} & = &
- 2xg_{\nu\be}R_\al\mbox{}^{\rh\la}\mbox{}_\mu +4x\de^\rh_\nu
R^\la\mbox{}_{\al\mu\be}+ (3x+y)g^{\rh\la}R_{\mu\al\nu\be}+
 2x \bar{\de}_{\nu\be ,}\mbox{}^{\rh\la}R_{\mu\al} -
(4x+2y)\de_\al^\rh R^\la\mbox{}_\mu g_{\nu\be} -
\nonumber \\
& - &
  2xg_{\nu\be}R_{\mu\al}g^{\rh\la}
+ \frac{y+2x}{2}\bar{\de}_{\mu\nu ,\al\be}R^{\rh\la}
- z g_{\nu\be}\bar{\de}_{\mu\al}\mbox{}^{\rh\la}R
+ \frac{1}{2}z\bar{\de}_{\mu\nu ,\al\be}g^{\rh\la}R
- 2z\bar{\de}_{\al\be}\mbox{}^{\rh\la}R_{\mu\nu}\, ;
\eeq
and
\beq
(\hat{W})_{\mu\nu ,\al\be} & = &
\frac{3x}{2}g_{\nu\be}R_{\mu\rh\la\si}R_{\al}\mbox{}^{\rh\la\si}
+\frac{x-y}{2}R^\rh\mbox{}_{\al\mu}\mbox{}^\la R_{\nu\be\rh\la} +
\frac{5x+y}{2} R^\la\mbox{}_{\al\mu}\mbox{}^\rh R_{\la\nu\be\rh}+
\frac{3x+y}{2}\, R^\la\mbox{}_\mu\mbox{}^\rh\mbox{}_\nu
R_{\rh\al\la\be}+
\nonumber \\
& + & \frac{y-5x}{2}\,R_{\mu\rh}R^\rh\mbox{}_{\al\nu\be}
+\frac{y+2x}{2}R_{\mu\al}R_{\nu\be}+
\frac{3y}{2}g_{\nu\be}R_{\mu\rh}R^\rh\mbox{}_\al
+\frac{3z}{2}g_{\nu\be}RR_{\al\mu}
-\frac{z}{2}RR_{\nu\be\al\mu} +
\nonumber \\
& + & zR_{\mu\nu}R_{\al\be} -
\frac{1}{4}\left( xR_{\rh\la\si\ta}^2+yR_{\rh\la}^2+zR^2\right)
\left( \bar{\de}_{\mu\nu\, ,\,\al\be}\right)\,.
\eeq
\end{widetext}
In the above formulas we used special condensed notations which
enable one to present the expressions in a relatively compact
way. The idea of these condensed notations is that all the
algebraic symmetries are implicit, including the symmetrizations
in the couples of indices $\,(\al\be)\leftrightarrow(\mu\nu)$,
$\,(\al\leftrightarrow\be)\,$ and
$\,(\mu\leftrightarrow\nu)\,$, and also
in the couple $\,(\rh\leftrightarrow\la)\,$ in the operator
$\hat{D}^{\rh\la}$. In order to obtain the complete formula
explicitly, one has to restore all the symmetries. For example,
\beq
R_{\mu\rh}R^\rh\mbox{}_{\al\nu\be}\to \frac{1}{2}
\left( R_{\mu\rh}R^\rh\mbox{}_{\al\nu\be}
+ R_{\al\rh}R^\rh\mbox{}_{\mu\be\nu} \right)
\nonumber
\eeq
restores the $\,(\al\be)\leftrightarrow(\mu\nu)$ symmetry.
The same procedure has to be applied also for the other
symmetries
$\,(\rh\leftrightarrow\la)$, $\,(\al\leftrightarrow\be)\,$
and $\,(\mu\leftrightarrow\nu)$.
\vskip 2mm

In order to use the Schwinger-DeWitt method for the fourth
derivative operator \cite{frts82}, we need to reduce it to the
minimal form (\ref{minimal high}). For this end one has to
multiply the operator (\ref{bilinear}) by the inverse matrix
$\,\hat{K}^{-1}$, given by
\beq
(\hat{K}^{-1})^{\mu\nu\,,\,\al\be}
= \frac{4}{y+4x} \,\bar{\de}^{\mu\nu\, ,\,\al\be}\,. \nonumber
\eeq
\vspace{1mm} Let us notice that the matrix $\,\hat{K}^{-1}\,$
is a $c$-number operator and hence this multiplication does not
affect the divergences. By straightforward algebra, one can find
the minimal operator
\beq
\hat{H}\,=\,\hat{K}^{-1}\,\hat{\cal H}
\,=\,\hat{1}\Box ^2 + \hat{V}^{\rh\la}\na_\rh\na_\la +\hat{U}\,,
\label{main}
\eeq where the new expressions
\beq
\hat{V}^{\rh\la}=\hat{K}^{-1}\hat{D}^{\rh\la} \qquad  \mbox{and}
\qquad \hat{U}=\hat{K}^{-1}\hat{W} \eeq already do not possess the
symmetry in $\,(\al\be)\leftrightarrow(\mu\nu)$. The expressions
for these two matrices are the following:
\begin{widetext}
\beq
(\hat{U})_{\mu\nu\, ,\,\al\be} & = & \frac{4}{y+4x}
\left\{\frac{3x}{2} g_{\nu\be}R_{\mu\rh\la\si}R_\al\mbox{}^{\rh\la\si}
+\frac{5x+y}{2}R^\la\mbox{}_{\al\mu}\mbox{}^\rh R_{\la\nu\be\rh} +
\frac{3x+y}{2}\,
R^\la\mbox{}_{\mu}\mbox{}^\rh\mbox{}_\nu R_{\rh\al\la\be}+
\frac{y-5x}{2}R_{\mu\rh}R^\rh\mbox{}_{\al\nu\be}+ \right.
\nonumber \\
& + & \left.\frac{y+2x}{2}R_{\mu\al}R_{\nu\be}+
\frac{3y}{2}g_{\nu\be}R^\la\mbox{}_\mu R_{\al\la}
- \frac{1}{4}\left(xR^2_{\rh\la\si\ta}+yR^2_{\rh\si}+zR^2\right)
\left(\bar{\de}_{\mu\nu\, ,\,\al\be}\right)+ \right.
\nonumber
\\
& + & \left. \frac{3z}{2}g_{\nu\be}RR_{\al\mu}
+\frac{x-y}{2}R^\rh\mbox{}_{\al\mu}\mbox{}^\la R_{\nu\be\rh\la}
+ zR_{\mu\nu}R_{\al\be}-\frac{z}{2}RR_{\nu\be\al\mu} \right\}\,,
\label{U}
\eeq
\end{widetext}
\beq
\hat{V}^{\rh\la} &=& \frac{4}{y+4x}\sum^{10}_{i=1} \, b_i\, {\bf k}_i\,,
\eeq
where
\beq
{\bf k}_1 & = & g_{\nu\be}\,g^{\rh\la}\,R_{\mu\al}\,;\quad \,
{\bf k}_2 = \bar{\de}_{\mu\nu\, ,\,\al\be}\,g^{\rh\la}\,R \,;
\eeq
\beq
{\bf k}_3 & = & g^{\rh\la}\,R_{\mu\al\nu\be}\,;\quad
{\bf k}_4 = \de_{\nu\be\, ,}\mbox{}^{\rh\la}\,R_{\mu\al}\,;
\nonumber \\
{\bf k}_5 & = & \de_{\nu\be\, ,}\mbox{}^{\rh\la}\,R\,g_{\mu\al}\,;\quad
{\bf k}_6 = \bar{\de}_{\mu\nu\, ,\,\al\be}\,R^{\rh\la}\, ;
\nonumber \\
{\bf k}_7 & = & \frac12(\,\de_\nu^{(\rh} \,R^{\la )}\mbox{}_{\al\be\mu}
+\de_\be^{(\rh} \,R^{\la )}\mbox{}_{\mu\nu\al}\, ) ;
\nonumber \\
{\bf k}_8 & = & g_{\nu\be}\,\de_{(\mu}^{(\rh} \,R^{\la )}\mbox{}_{\al )}\, ;\quad
{\bf k}_9  =  g_{\nu\be}\,R_{(\al}\mbox{}^{\rh\la}\mbox{}_{\mu )}\, ;
\nonumber \\
{\bf k}_{10} & = & \frac12\,(\,\bar{\de}_{\al\be\, ,}\mbox{}^{\rh\la}\,R_{\mu\nu}
+\bar{\de}_{\mu\nu\, ,}\mbox{}^{\rh\la}\,R_{\al\be}\, )
\nonumber
\eeq
and
\beq
b_1 &=& -2x\, ;\qquad b_2 = z/2\, ;\qquad
b_3 = 3x+y\, ;
\nonumber \\
b_4 & = & 2x\, \qquad b_5 = -z\,\qquad
b_6 = x+y/2\,;
\nonumber \\
b_7 & = & -4x\, ;\;\; b_8 = -4x-2y\, ;\;\;
b_9 = -2x\, ;
\nonumber \\
 b_{10} & = & -2z\,.
\label{b}
\eeq
The above form of $\,\hat{V}^{\rh\la}\,$ is helpful in organizing
the cumbersome calculations of divergences which will be described
in the next section.

\section{\label{4}Derivation of divergences}

The algorithm for the 1-loop divergences corresponding to the
minimal fourth order operator can be written as \cite{frts82}
(here we use the Euclidean signature of the metric, in order
to be consistent with \cite{frts82})
\beq
\frac12 \ln \Det \hat{H}\Big|_{\rm div} & = &
-\,\frac{\mu^{n-4}}{(4\pi )^2(n-4)}\, \times
\nonumber \\
& \times & \int d^nx\sqrt{g}\,
\tr \lim_{x^\prime \to x} a_2(x^\prime ,x)\,,
\label{a2}
\eeq
where
\beq
\lim_{x^\prime \to x}\,a_2(x^\prime\,,\,x)
& = & {\frac {\hat{1}}{90}}
\,R_{\mu\nu\al\be}^2-{\frac {\hat{1}}{90}}\,R_{\mu\nu}^2+
\frac{\hat{1}}{36}\,R^2 \,+\,
\nonumber
\\
& + & \frac{1}{6}\hat{\cal{R}}_{\mu\nu}\hat{\cal{R}}^{\mu\nu}
-\hat{U}+
\frac{1}{12}R\hat{V}^\rh\mbox{}_\rh -
\label{gr-formula}
\\
& - & \frac{1}{6}R_{\rh\la}\hat{V}^{\rh\la} +
\frac{1}{48}\hat{V}^\rh\mbox{}_\rh\, \hat{V}^\la\mbox{}_\la +
\frac{1}{24}\hat{V}_{\rh\la}\hat{V}^{\rh\la}\,. \nonumber \eeq
Here $\hat{\cal{R}}_{\mu\nu}$ is the commutator of the covariant
derivatives acting in the tensor $\,h^{\al\be}\,$ space, \beq
\hat{\cal{R}}_{\mu\nu}\,=\,[\na_\mu\,,\,\na_\nu ]\,.
\label{commute}
\eeq
The full collection of the traces of the
expressions (\ref{gr-formula}) is presented in the Appendix
\ref{B}.

It is straightforward to find the contributions of the weight
and ghost operators (the algorithm for the non-minimal vector
operator can be found in \cite{frts82,avba})
\begin{widetext}
\beq
-\frac{i}{2}\,{\rm ln\; det}\, \hat{Y}|_{{\rm div}} & = &
-\frac{1}{(n-4)(4\pi )^2}\int d^4 x\sqrt{-g}\,\left\{\,
{\frac {11}{180}}\, R_{\mu\nu\al\be}^2-
{\frac {43}{90}}\, R_{\mu\nu}^2+\frac{1}{9}\,R^2\,\right\}
\label{-we}
\\ \nonumber \\
-i\; {\rm ln\; det}\, \hat{\cal{H}}_{gh}\,|_{{\rm div}}
& = &
-\frac{1}{(n-4)(4\pi )^2}\int d^4 x\sqrt{-g}\,\left\{\,
{\frac {11}{90}}\,E-\left (\frac{1}{3}\,{\xi}^{2}
-\frac{4}{3}\,\xi+{\frac {7}{15}}
\right ) R_{\mu\nu}^2 -
\left (\frac{1}{6}\,{\xi}^{2}-\frac{1}{3}\,\xi
+{\frac {17}{30}}\right ) R^2\,\right\} , \nonumber \\
\label{-gh}
\eeq
\end{widetext}
where the parameter $\xi$ is given by
\beq
\xi = \frac {n-2}{2(n-1)}\, .
\eeq

Collecting all the results from (\ref{gr-formula}),
(\ref{-we}) and (\ref{-gh}) according to (\ref{e}) and
using the formulas from Appendix B, we
arrive at the functional trace of the overall coincidence
limit of the $\,a_2(x^\prime,x)$-coefficient
\beq
A^t_2 & = & \lim_{x^\prime\to x}\mbox{sTr}\,a^t_2(x^\prime,x)
\,=\,\lim_{x^\prime\to x}\,\left[\Tr a_2(x^\prime,x)(\hat{\cal H})
- \right.\nonumber
\\
& - & \left. \Tr a_2(x^\prime,x)(\hat{Y})
-2\Tr a_2(x^\prime,x)(\hat{\cal H}_{gh})\right]\,.
\label{over}
\eeq
The last expression can be regarded as a functional supertrace
of the coincidence limit of the $\,a_2(x^\prime,x)$-coefficient
of the differential operator acting in the direct product
of the tensor $\,h_{\mu\nu}$, vector (third ghost) and vector
ghost spaces. The sign difference between the different terms
in (\ref{over}) is due to the different Grassmann parity of
the fields, and the operator $\,\Tr\,$ includes integration,
as usual.

Let us present the result in terms
of the parameters $\,\eta\,$ and $\,\la\,$:
\beq
A_2^t & = & -\mu^{n-4}\,\int d^nx\sqrt{-g}\,\left\{\,
\be_1(n)\, E+ \right. \nonumber
\\
& + & \left. \be_2(n)\, C^2+\be_3(n)\,R^2\,\right\}\, ,
\label{finaldiv}
\eeq
where the coefficients ($\be$-functions) $\,\be_1(n)$,
$\,\be_2(n)\,$ and $\,\be_3(n)\,$ are given by the
expressions
\beq
\be_i(n) = \de^{(0)}_i +\de^{(1)}_i\,\eta +\de^{(2)}_i\,\eta^2\,,
\;\; i=(1,2,3)\,.
\eeq
The coefficients $\,\de^{(i)}_j$\, are the following functions:
\begin{widetext}
\beq
\de^{(0)}_1 & = &
-\,{\frac {15\,{n}^{6}+86\,{n}^{5}+201\,{n}^{4}-4842
\,{n}^{3}+8104\,{n}^{2}+6624\,n-9648}{2880\,\left (n-1\right )
\left (n-3\right )^{2}}}\, ,
\nonumber
\\
\nonumber
\\
\de^{(1)}_1 & = &
-\,{\frac {\left(n-4\right)\,\left (n-2\right )
\left ({n}^{5}-8\,{n}^{4}+39\,{n}^{3}-40\,{n}^{2}
-196\,n+192\right )\lambda}{48\,n\left (n-1 \right )
\left (n-3\right )^{2}}}\, ,
\nonumber
\\
\nonumber
\\
\de^{(2)}_1 & = &
-\,{\frac {\left (n-4\right )^{2}\,
\left ({n}^{3}+9\,{n}^{2}+14\,n+12\right)
\left (n-2 \right )^{2}\,{\lambda}^{2}}
{48\,\left (n-3\right )^{2}{n}^{2}}}\, ,
\nonumber
\\
\nonumber
\\
\de^{(0)}_2 & = &
{\frac {\left (n-2\right )
\left (5\,{n}^{6}+299\,{n}^{5}-1162\,{n}^{4}-2570\,{n}^{3}
+15056\,{n}^{2}-18528\,n+6720\right )}
{960\,n\left (n-1\right )\left (n-3\right )^{2}}}\, ,
\nonumber
\\
\nonumber
\\
\de^{(1)}_2 & = &
{\frac {\left (n-4\right )
\left ({n}^{4}-3\,{n}^{3}+50\,n-36\right )
\left (n-2\right )^{2}\lambda}{48\,n\left (n-1\right)
\left (n-3 \right )^{2}}}\, ,
\nonumber
\\
\nonumber
\\
\de^{(2)}_2 & = &
{\frac {\left (n-4\right )\left ({n}^{3}+10\,{n}^{2}+10\,n+24
\right )\left (n-2\right )^{3}{\lambda}^{2}}
{48\,\left (n-3\right )^{2}{n}^{2}}}\, ,
\nonumber
\\
\nonumber
\\
\de^{(0)}_3 & = &
{\frac {\left (n-4\right )
\left (5\,{n}^{5}+22\,{n}^{4}+179\,{n}^{3}-930\,{n}^{2}-112\,n+816
\right)}{960\,\left (n-1\right )^{2}\left (n-3\right )}}\, ,
\nonumber
\\
\nonumber
\\
\de^{(1)}_3 & = &
{\frac {\left (n-4\right )\left ({n}^{4}-4\,{n}^{3}-{n}^{2}+10\,
n-12\right )
\left (n-2\right )^{2}\lambda}{24\,n\left (n-1\right )^{2}
\left (n-3\right )}}\, ,
\nonumber
\\
\nonumber
\\
\de^{(2)}_3 & = &
{\frac {\left (n-4\right )^2 \left(n+1\right)
\left ({n}^{2}+2\,n+12\right)
\left (n-2\right )^3\,{\lambda}^2}{96\,\left (n -1\right )
\left (n-3\right ){n}^{2}}}\, .
\label{coefficients}
\eeq
\end{widetext}

The above coefficients, despite their chaotic appearance,
provide  a lot of important information. First of all,
they show that for $\,n\neq 4\,$ the Gauss-Bonnet topological
term contributes to the
effective action in a non-trivial way, and in particular
produces the $\,\int\sqrt{-g}R^2$-type term. On the other
hand, it is
remarkable that all $\,\de^{(1)}_i\,$ and $\,\de^{(2)}_i\,$
coefficients are proportional to $\,(n-4)$. Hence,
for $\,n=4\,$ one can see that the $\,\eta$-dependence
completely disappears, and the
final result, Eq. (\ref{finaldiv}) becomes very simple.
Let us write down the expression for the one-loop
divergences
\beq
\Ga ^{(1)}_{{\rm div}} &=& -
\frac{\mu^{n-4}}{(4\pi )^2(n-4)}\int d^n x\sqrt{-g}
\left\{\frac{137}{60}E+\frac{199}{15} W \right\}
\nonumber
\\
&=&
\frac{\mu^{n-4}}{(4\pi )^2(n-4)}\int d^n x\sqrt{-g}
\left\{\frac{87}{20} E - \frac{199}{30}C^2\right\}
\nonumber
\\
\label{antoniadis}
\eeq
where we used Eq. (\ref{W}) and the pseudo-Euclidean
signature. The expression (\ref{antoniadis})
coincides with the one derived by Antoniadis, Mazur and
Mottola in \cite{amm}. Both coefficients in (\ref{antoniadis})
also coincide with the ones derived by Fradkin and Tseytlin
in \cite{frts82}. However, we do not meet the polemical
$\,\int\sqrt{-g}R^2$-type divergence \cite{frts82} and
hence there is no need to apply the conformal regularization
\cite{truffin,frvi} discussed in the Introduction.
As far as our calculation is
seriously tested by the cancellation of the numerous
$\,\eta$-dependent terms in the $\,n\to 4\,$
limit, we strongly believe in its correctness. Thus, the
$\,\int\sqrt{-g}R^2$-type one-loop divergence does not
show up in the one-loop effective action of
Weyl quantum gravity.

The expression (\ref{antoniadis}) does not contain the
divergences in the cosmological constant and of the
Einstein-Hilbert type. This fact is due to our choice
of the regularization procedure. As far as we start
from the conformally invariant action, the theory does
not possess dimensional parameters and therefore the
divergences of these two sorts may be only quartic
and quadratic ones. Of course, in the dimensional
regularization which we are using here, the quartic
and quadratic divergences do not show up. However, one
can easily see the cosmological and linear in $R$
divergences in other regularization schemes, for
example in the covariant cut-off \cite{frts82}
or in the covariant Pauli-Villars \cite{anomayo}
regularizations. Of course,
the logarithmic divergences which we have
calculated (\ref{antoniadis}) do not depend on the
choice of the regularization scheme.

Including the matter fields one meets additional
contributions to the divergent
coefficients in (\ref{antoniadis}). As it was already
noticed in the Introduction, the conventional
scalars, fermions and vectors give contributions
of the same sign to both $\,\be_1\,$ and $\,\be_2\,$,
while the contributions of higher derivative scalar
and fermion have opposite sign \cite{rei,frts-sugra,cofe}.
The sign of the coefficients in (\ref{antoniadis})
coincides with the one of the conventional fields.
Hence, since
the $\,\int R^2\,$-type divergence is absent, one can
use the method of \cite{cofe} and adjust the number
of higher derivative scalars and fermions such
that the one-loop divergences completely cancel.
In a more complicated
situation, when the matter coupled to quantum
gravity possesses self-interaction, the quantum
gravitational effects modify the divergences and the
renormalization group trajectories, also, in the matter
sector of the theory. This issue has been studied
in details \cite{bush86,BKSVW,shap} for the case of the
higher derivative gravity \footnote{Let us remark
that the quantum gravity theory based on General
Relativity does not admit this sort of investigation
because the corresponding theory is not renormalizable.}.
In both conformal \cite{bush86} and general
\cite{BKSVW,shap} cases the effect of quantum
gravity is rather smooth and always favors the
asymptotic freedom in the matter fields sector.
Due to the absence of the $\,\int R^2\,$-type divergence
the investigation for the conformal case can be indeed
performed without the special conformal regularization
(which we discussed in the Introduction), while the
quantum-gravitational corrections to the $\be$-functions
in the matter sector are exactly the ones derived in
\cite{bush86} (see also \cite{book}). The reason is
that these $\be$-functions do not depend on the
scalar curvature and hence are not affected by the
conformal regularization.

\section{\label{5}Renormalization group equations}

The renormalization group (RG) equations for the theory
(\ref{Weyl}) may be considered in two different
ways \footnote{The
renormalization group equations for the higher derivative
quantum
gravity where introduced in \cite{julve,salam,frts82}.
The renormalization group equations for the higher
derivative gravity coupled to GUT's have been
derived in \cite{bush86,shap} for the conformal case
and in \cite{frts82,BKSVW,shap} for the general case.}.
The first possibility is to take usual $\,n=4\,$
$\,\,\be$-functions, in this case we meet exactly the
same RG equations as in \cite{frts82}. It proves useful
to introduce a new
parameter $\,\rho=-1/\eta$. Let us remark that the
choice of $\,\la\,$ as a coupling constant in the
action (\ref{Weyl}) is fixed, because {\it (i)}  $\,\la\,$
is a parameter of the loop expansion in this theory;
{\it (ii)} One can not change the sign of $\,\la\,$
without changing the positivity of the graviton energy.
At the same time there are no similar constraints for
the coefficient of the Gauss-Bonnet term and therefore
the choice can be made according to convenience.
The usual $\,n=4\,$ renormalization  group equations for
$\,\la\,$ and $\,\rho\,$ have the
form
\beq
\frac{d\la}{dt} &=& \mu\frac{d\la}{d\mu}\Big|_{n=4}
=\be_\la(4)=-a^2\,\la^2\,,\; \la(0)=\la_0\,;
\nonumber
\\
\nonumber
\\
\frac{d\rho}{dt}&=& \mu\frac{d\rho}{d\mu}\Big|_{n=4}
=\be_\rho(4)=-b^2\,\rho^2\,,\; \rho(0)=\rho_0\,,
\label{r g}
\eeq
where  \beq
a^2\,=\,\frac{199}{15\,(4\pi)^2}\,,
\qquad\qquad
b^2\,=\,\frac{261}{60\,(4\pi)^2}\,.
\label{a b}
\eeq
The above equations indicate the UV asymptotic freedom
in both parameters. In other words, there is a single
fixed point $\,\la=\rho=0\,$ and it is
stable in the high energy limit $\,t\to \infty$.

Let us now consider a more complicated version of the
renormalization group equations, taking the dimension
$\,n=4-\ep\,$ \ for $\,\,-1 \leq \ep < 1$. Mathematically
this means that we do not
take the limit $\,n\to 4\,$ in the equations (\ref{r g}).
The renormalization group equations which emerge as a
result of this procedure will be different from (\ref{r g})
and one can expect to see qualitative effects of the
Gauss-Bonnet term in this framework.

Similar approach to the renormalization group proved
fruitful in the two-dimensional quantum gravity
\cite{GKT}, due to its relation to the concept of
asymptotic safety \cite{Weinberg 2-e} and to the discussion
of the universality classes of quantum gravity theories
\cite{KN}. The main idea of $\,2-\ep\,$ quantum
gravity is the following. In the precisely $\,n=2\,$
dimensions, quantum gravity is a
topological theory similar to the one which we meet
in $\,n=4\,$ starting from the pure Gauss-Bonnet term.
But, if we generalize the theory for
$\,n=2-\ep\,$, there is a dynamics (different from the
Gauss-Bonnet theory, where the propagator does not
appear even for $\,n\neq 4$) and at the quantum level
one meets a non-trivial UV fixed point of the
renormalization group \cite{GKT,Weinberg 2-e,KN}.
Keeping this example in
mind, it is natural to expect that the effect of the
Gauss-Bonnet term on the renormalization group equations
in $\,n=4-\ep\,$ may be non-trivial and in particular
may produce new fixed points.

Consider the renormalization group equations for
$\,\la\,$ and $\,\rho\,$ in $\,n=4-\ep\,$ dimension.
The naive form of the $\,\be$-functions would be based
on the ``standard'' expressions (\ref{r g})
\beq
\be_\la\,=\,-\ep\la+\be_\la(4)\,,\qquad
\be_\rho\,=\,-\ep\rho+\be_\rho(4)\,,
\label{naive}
\eeq
indicating the one extra nonzero fixed point for
each of the effective charges $\,\la(t)\,$ and $\,\rho(t)$.
Indeed, the fixed point $\,\la=\rho=0\,$ remains
stable in UV for $\,\ep > 0$.
However, this naive consideration is incorrect because
the Gauss-Bonnet term gets dynamical in $\,n\neq 4$,
affecting the renormalization group equations in a
non-trivial way. Using the
expressions (\ref{coefficients}), we arrive at the
following correct form of the renormalization group
equations, quite different from (\ref{naive}):
\beq
\frac{d\rho}{dt} & = & -\,\ep\rho\,+\,\frac{1}{(4\pi)^2}
\,\Big(\,f_1\rho^2 - f_2\,\la\rho +f_3\,\la^2\,\Big)\,,
\nonumber
\\
\nonumber
\\
\frac{d\la}{dt} & = & -\,\ep\la\,-\,\frac{2\la^2}{(4\pi)^2}
\,\Big(\,g_1 - g_2\,\frac{\la}{\rho} +
g_3\,\frac{\la^2}{\rho^2}\,\Big)\,.
\label{beta 2}
\eeq
The coefficients $\,f_{1,2,3},\,\,g_{1,2,3}\,$
may be expressed via the coefficients $\,\de^{(i)}_j$
from (\ref{coefficients}) as
\beq
f_1 = \de_1^{(0)}\,,\qquad
f_2 = \de_1^{(1)}/\la\,,\qquad
f_3 = \de_1^{(2)}/\la^2\,,
\nonumber
\\
\nonumber
\\
g_1 = \de_2^{(0)}       \,,\qquad
g_2 = \de_2^{(1)}/\la   \,,\qquad
g_3 = \de_2^{(2)}/\la^2 \,.
\label{f g}
\eeq
One can remark that $\,\,f_{1,2,3}\,\,$ and
$\,\,g_{1,2,3}\,\,$
depend only on the parameter $\,\ep\,$ and not on the
couplings. In the limit $\,\ep=0\,$ we come back to the
equations (\ref{r g}).
The renormalization group equations (\ref{beta 2}) are
non-linear and do not admit a simple analytic solution.
For this reason we shall start from the
search of the fixed points, that are the values of
$\,\la\,$ and $\,\rho\,$ for which both $\,\be$-functions
vanish. Consequently, we explore the stability of these
fixed points and establish the renormalization group
flows for some particular values of $\,\ep$.

In order to find fixed points, we consider
the particular values of the parameter,
$\,\ep=0.9\,$, $\,\ep=0.1\,$, $\,\ep=0.01$, $\,\ep=-0.01\,$,
$\,\ep=-0.1$ and $\,\ep=-1$. The numerical values
of the coefficients for these cases are presented in the
Table 1. The point $\ep =0.9$ is numerically close to
$\ep = 1$ ($n=3$) where the
expressions for the $\be$-functions become singular.



\begin{center}
\begin{tabular}{|l|l|l|l|l|l|l|}
\hline
$\ep$ & $f_1$ & $f_2$ & $f_3$ & $g_1$ & $g_2$ & $g_3$    \\
\hline
0.9 & -16.77 & -28.73 & -36.48 & 2.359 & -42.51 & -46.98 \\
\hline
0.1  & -4.301 & 0.08 & -0.016  & 6.385 & -0.174 & -0.318 \\
\hline
0.01 & -4.344 & 0.008 & -0.0001 & 6.608 & -0.016 & -0.03 \\
\hline
-0.01 & -4.356 & -0.008 & -0.0001 & 6.659 & 0.016 & 0.03 \\
\hline
-0.1 & -4.416 & -0.086 & -0.013 & 6.902 & 0.146 & 0.286 \\
\hline
-1 & -5.416 & -0.947 & -0.81 & 9.98 & 1.087 & 2.526
\\
\hline 
\end{tabular}
\end{center}
\begin{quotation}
{\large\sl Table 1.} Numerical values of the coefficients
for the particular values of $\,\ep$.
\end{quotation}

The numerical analysis shows that for each of the choices
$\,\ep=0.9\,$, $\,\ep=0.1\,$ and $\,\ep=0.01\,$ there are four fixed
points which are new compared to the $\,\ep=0\,$ case; while
for the values $\ep =-0.01\, ,\, -0.1\, ,\, -1$, there are two new fixed
points. The values of the parameters corresponding to these
fixed points are shown in the Table 2.
\vskip 5mm

\begin{center}
\begin{tabular}{|l|l|l|l|l|}
\hline 
Fixed Point
for $\ep=0.9$    &  1      &  2      &  3     &  4      \\
\hline
$\la_i$           & 0       & -6.817 & -5.945 & 1.807 \\
\hline
$\rho_i$         & -8.475  & -10.75 & -12.52 &  -3.05   \\
\hline \hline
Fixed Point
for $\ep=0.1$    &  1      &  2      &  3     &  4      \\
\hline
$\la_i$           & 0       & -14.421 & -1.232 & 16.236 \\
\hline
$\rho_i$ & -3.671 & -3.159  & -3.647 & -3.706           \\
\hline \hline
Fixed Point
for $\ep=0.01$    &  1      &  2      &  3     &  4     \\
\hline
$\la_i$           & 0       & -5.228 & -0.119 & 5.457 \\
\hline
$\rho_i$ & -0.364 & -0.351  & -0.363 & -0.371  \\
\hline \hline
Fixed Point
for $\ep=-0.01$    &  1      &  2      &  3     &  4      \\
\hline
$\la_i$           & 0       & $\;\;$ --- & $\;\;$ --- & 0.1186 \\
\hline
$\rho_i$       & 0.3625  & $\;\;$ ---  & $\;\;$ ---   & 0.3628 \\
\hline \hline
Fixed Point
for $\ep=-0.1$    &  1      &  2      &  3     &  4     \\
\hline
$\la_i$           & 0       & $\;\;$ --- & $\;\;$ --- & 1.147 \\
\hline
$\rho_i$ & 3.576 & $\;\;$ ---  & $\;\;$ --- & 3.597
\\
\hline \hline
Fixed Point
for $\ep=-1$    &  1      &  2      &  3     &  4     \\
\hline
$\la_i$           & 0       & $\;\;$ --- & $\;\;$ --- & 8.001 \\
\hline
$\rho_i$ & 29.157 & $\;\;$ ---  & $\;\;$ --- & 30.239
\\
\hline
\end{tabular}
\end{center}
\begin{quotation}
{\large\sl Table 2.} Numerical values of the parameters
corresponding to the new fixed points. No one of them has
analog in the $\,\ep=0\,$ case.
\end{quotation}
\vskip 2mm

The stability properties of the fixed points
can be easily investigated in the linear approximation. The
result is that, in the cases $\,\ep=0.9\,$, $\,\ep=0.1\,$
and $\,\ep=0.01$,
the fixed points $\,(\la_1,\rho_1)\,$ and $\,(\la_2,\rho_2)\,$
are saddle points while the fixed points
$\,(\la_3,\rho_3)\,$ and $\,(\la_4,\rho_4)\,$ are absolutely
unstable in the UV limit $\,t\to\infty$. It is worth noticing
the contrast with the $\,\ep=0\,$ renormalization group equations
(\ref{r g}) with a single UV stable fixed point $\,\la=\rho=0$.
In the $n > 4$ cases $\,\ep=-0.01\,$, $\,\ep=-0.1\,$
and $\,\ep=-1\,$
there are only
two extra fixed points, one of them UV-stable and IR-unstable
and another one a saddle point (unstable in both UV and IR regimes).
 As shown
in the Figures 1 and 2, there are no additional (compared to the
standard $\ep =0$
case) UV-stable fixed points for positive $\ep$, and at the same
 time, for negative
$\ep$ there is always one additional fixed point with stability
 in the UV domain.

\begin{figure*}
$$
\begin{array}{l}
\psfig{file=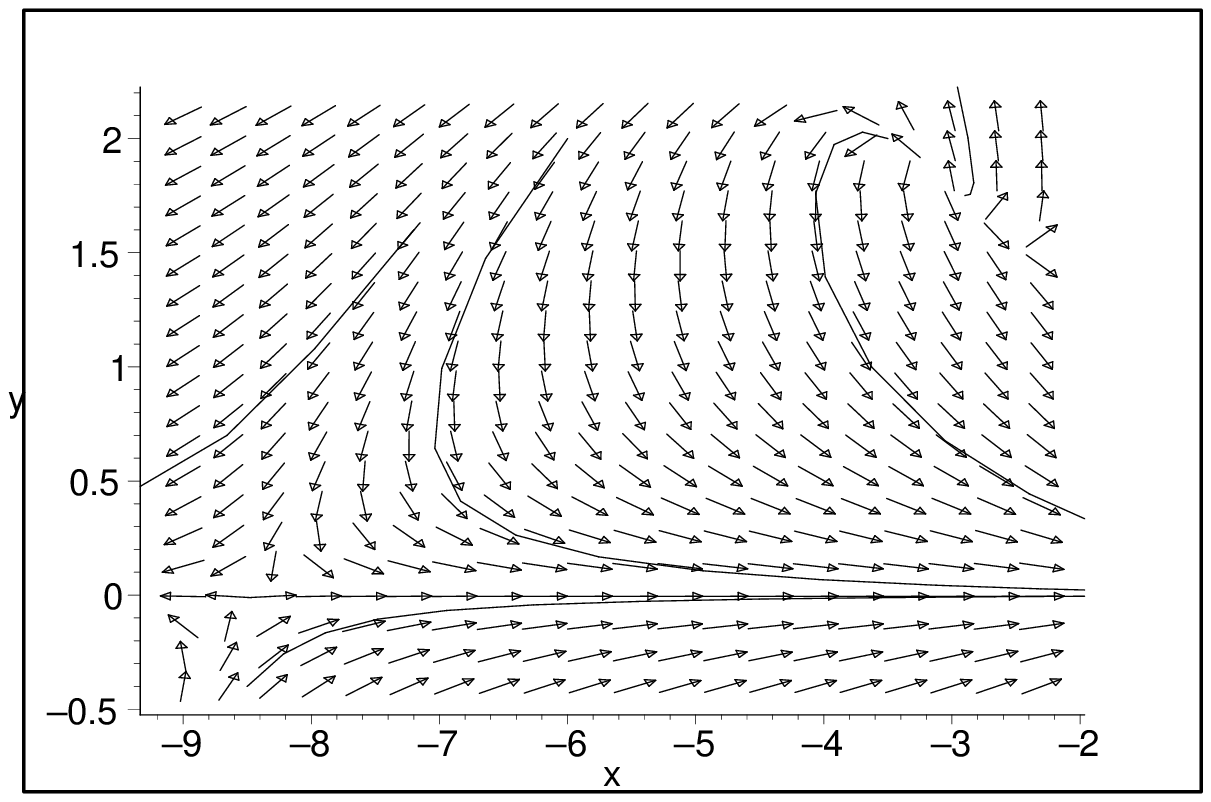,width=8cm,height=5.5cm}\;\;
\psfig{file=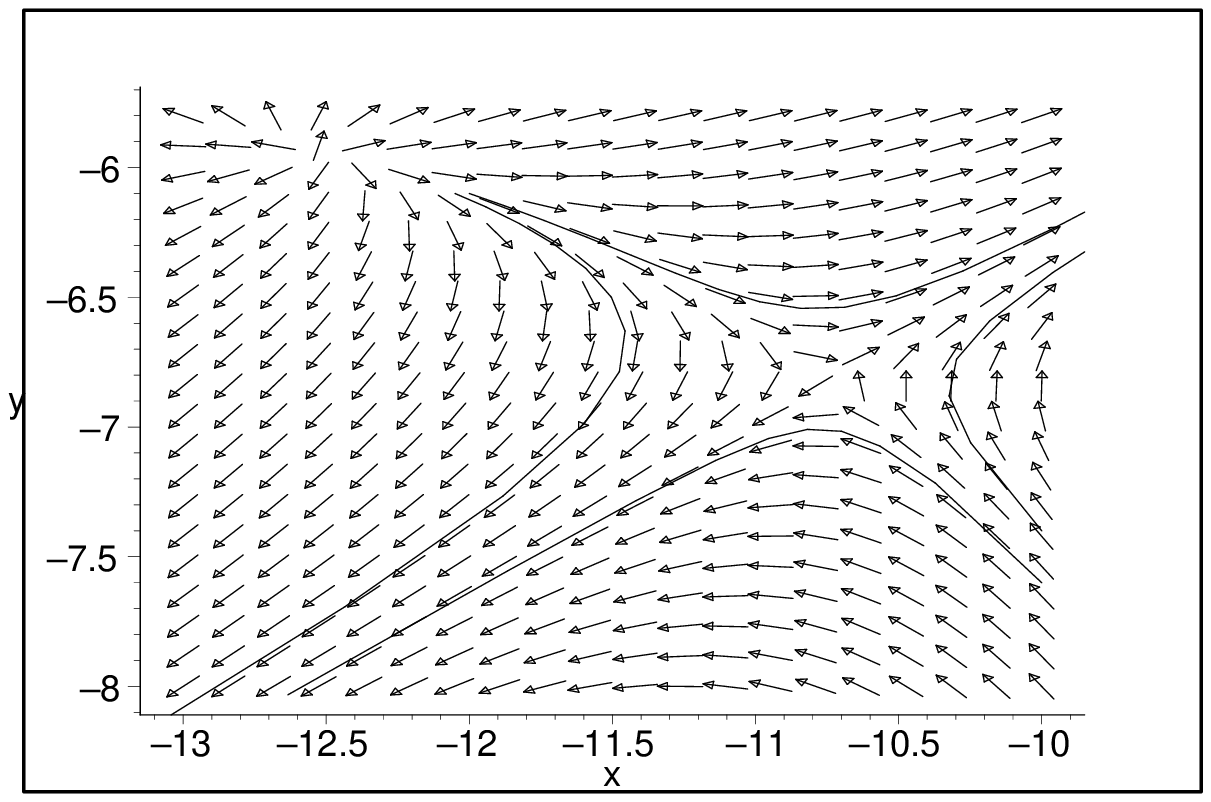,width=8cm,height=5.5cm}
\end{array}
$$
\vspace*{8pt} \caption{Diagrams for $\ep = 0.9$. The $y$-axis
represents the coupling $\la$ and the $x$-axis, the coupling
$\rho$, as well as in all subsequent plottings. The left diagram
shows the fixed points 1 and 4, and the other shows the points 2
and 3. The labels of the fixed points correspond
to the numeration in Table 2.
The arrows indicate the direction of the renormalization
group trajectory at the given point (we have also drawn some
trajectories for illustrative purpose). One can distinguish
stable, completely unstable and saddle fixed points at these and
further diagrams. } 
\end{figure*}

\begin{figure*}
$$
\begin{array}{l}
\psfig{file=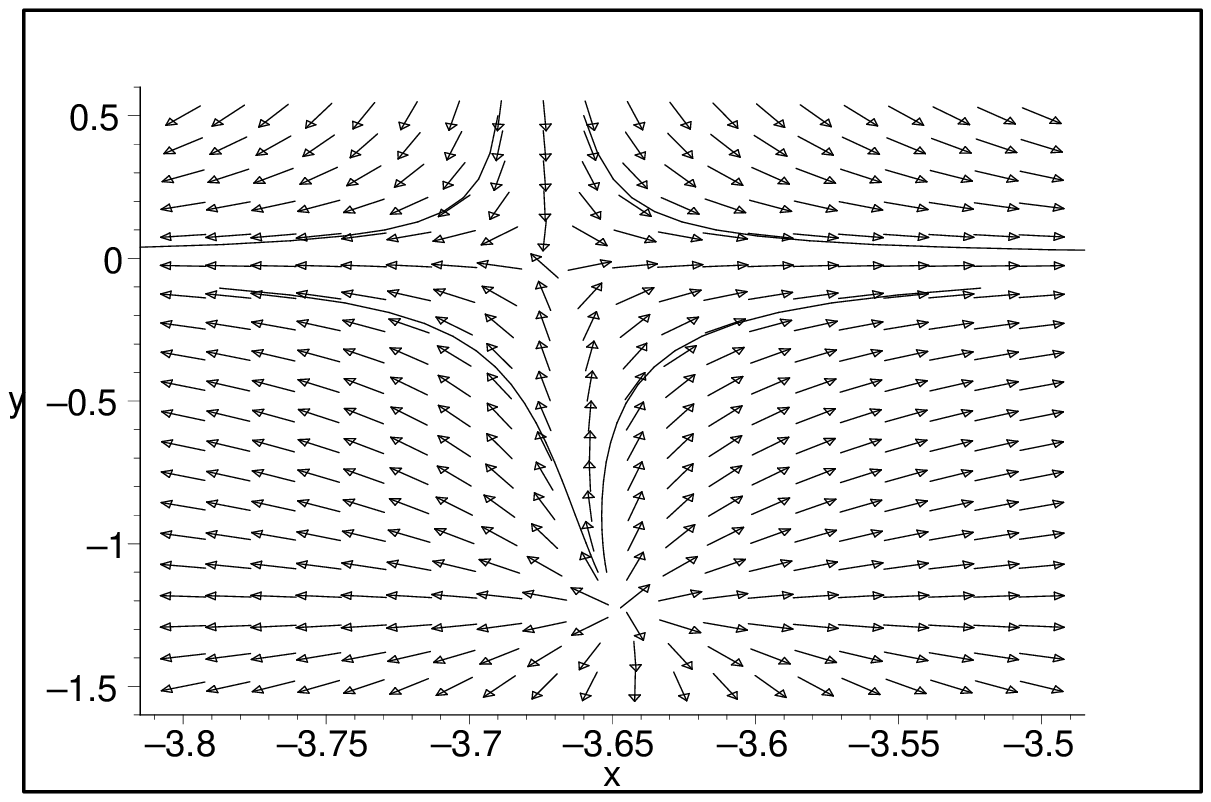,width=8cm,height=5.5cm}\;\;
\psfig{file=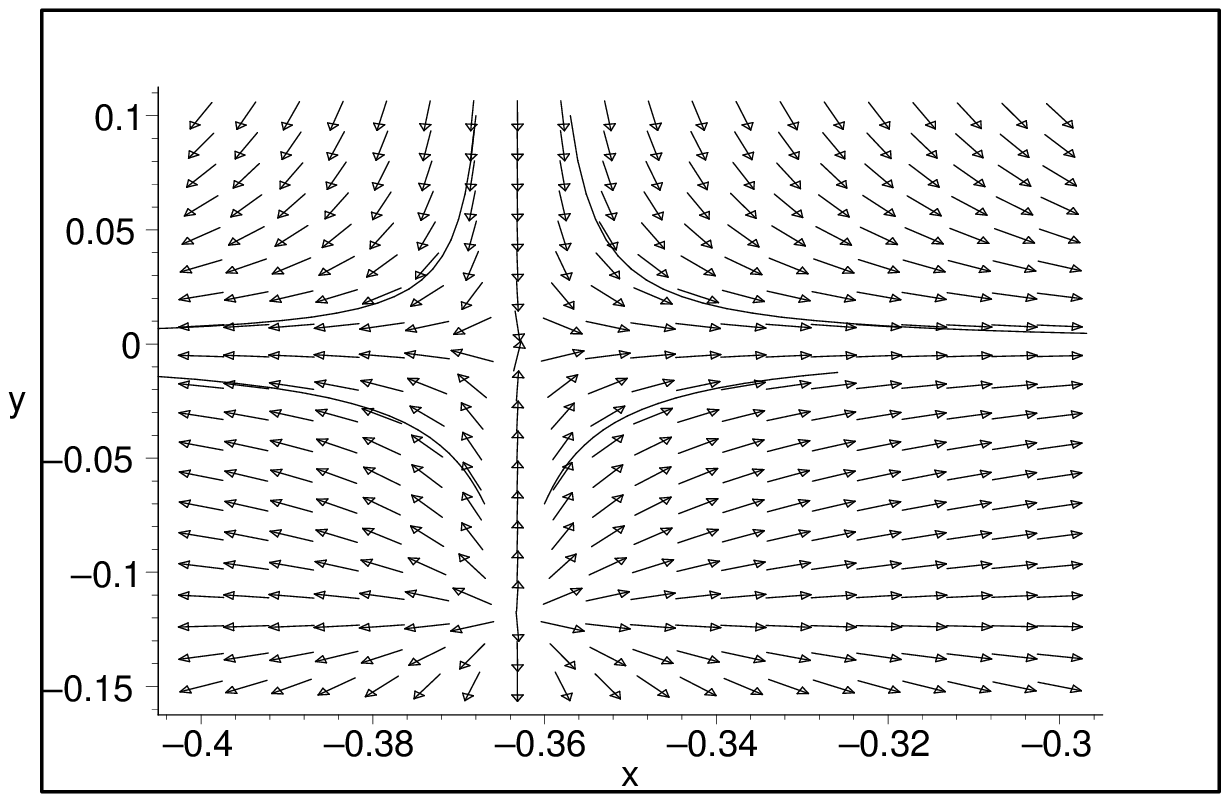,width=8cm,height=5.5cm}
\end{array}
$$
\vspace*{8pt} \caption{The fixed points 1 and 3 are shown for the
cases $\ep = 0.1$ (left) and $\ep = 0.01$ (right). Clearly, the
point 1 is a saddle point (unstable) and the point 3 is
UV-unstable. The points 2 and 4 are similar (saddle and
UV-unstable, respectively) and are not plotted.} 
\end{figure*}


\begin{figure*}
$$
\begin{array}{l}
\psfig{file=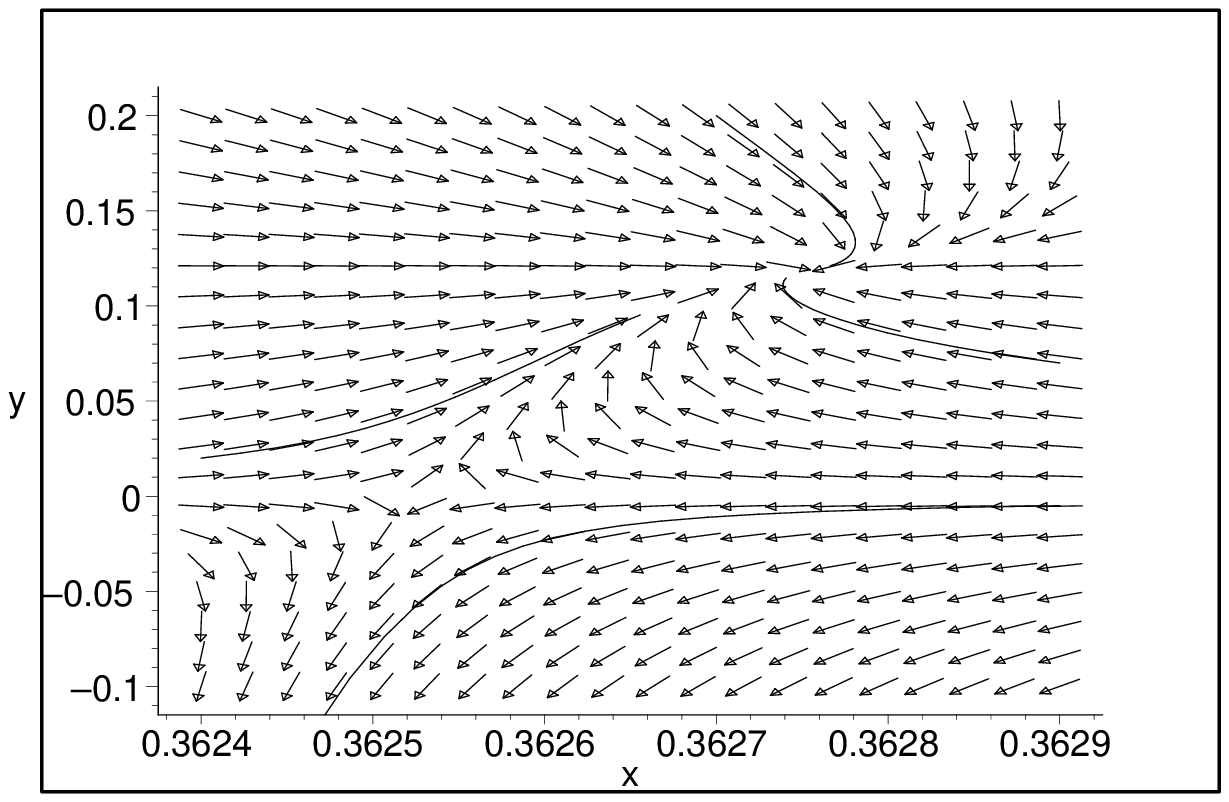,width=8cm,height=5.5cm}\;\;
\psfig{file=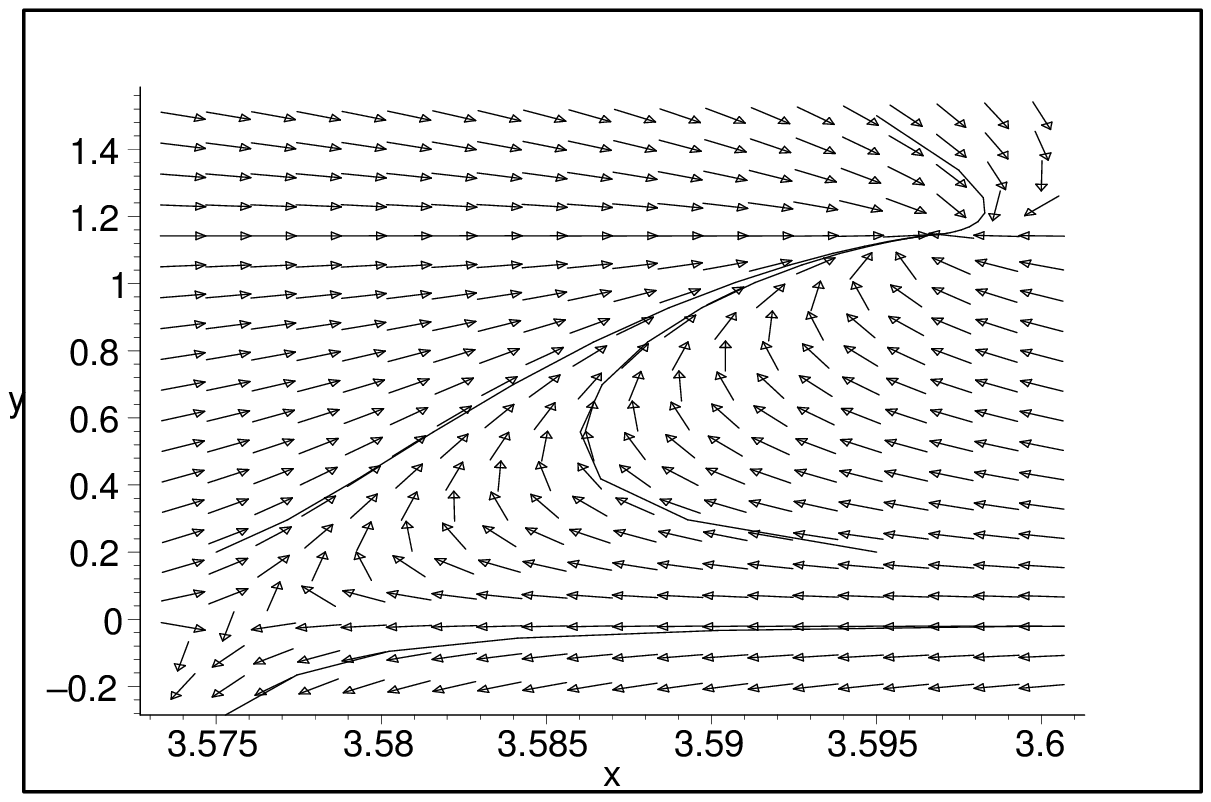,width=8cm,height=5.5cm}
\end{array}
$$
\vspace*{8pt}
\caption{Fixed points 1 and 4 for the cases $\ep = -0.01$
(left) and $\ep = -0.1$ (right). The point 1 is a saddle point
(unstable) and the point 4 is UV-stable, contrary to the
analogous points with positive $\ep$.}
\end{figure*}


\begin{figure}
\centerline{\psfig{file=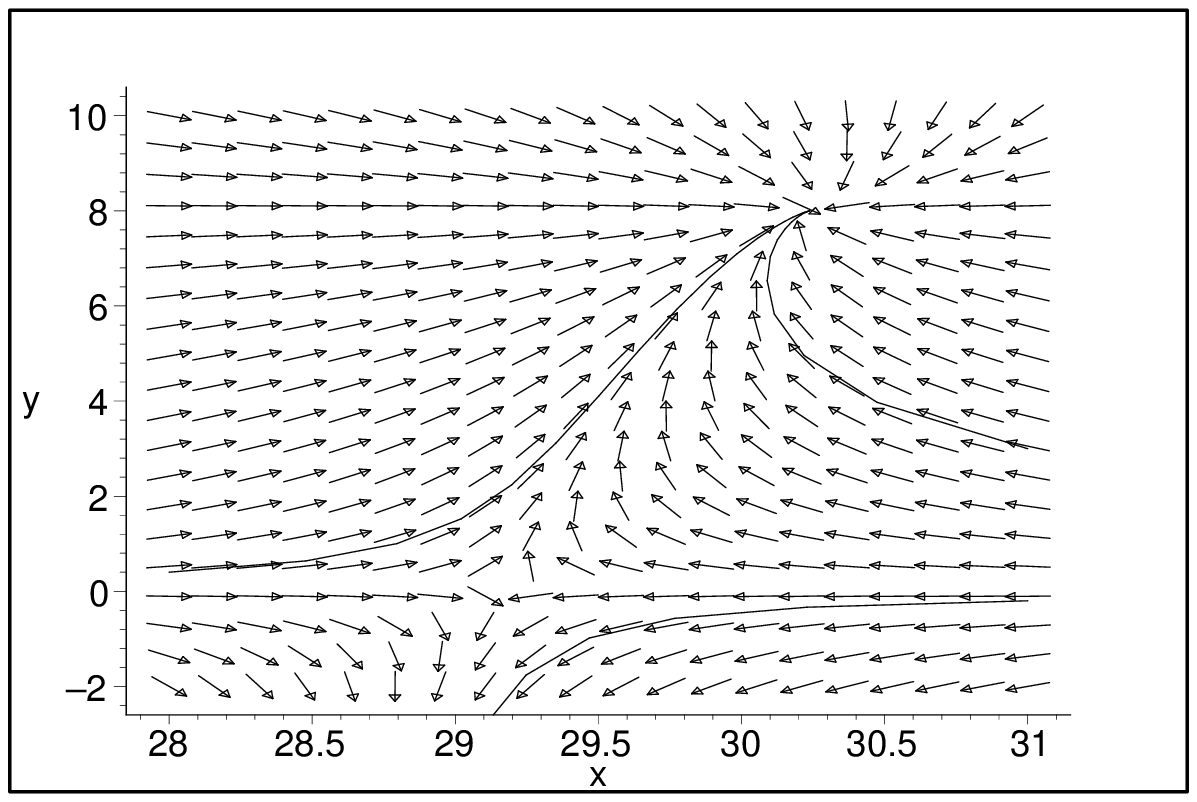,width=8cm,height=5.5cm}}
\vspace*{8pt} \caption{The case $\ep = -1$, with a saddle point
(1) and a UV-stable one (4).}
\end{figure}


An interesting observation concerning the renormalization group
trajectories is that no one of them crosses the line $\,\la=0$. In
other words, the renormalization group flow in this theory is
divided into two separate parts: one with $\,\la > 0\,$
corresponds to the positively defined energy of the gravitons
(massless spin-2 mode) and another one to the unphysical graviton
sector with $\,\la < 0$. There are examples of the qualitatively
new UV-stable fixed points with $\,\la > 0\,$ (see Figures 3,4).
At the same time there are no such examples for the case $\,\la <
0$. One can suppose that this property of the fixed points is
related to the limit $\ep\to 0$, where all new UV-stable fixed
points presumably should tend to $\,\la = 0$.

It is obvious that no one of the fixed points which we have found
so far coincides with the standard one $\,\la=\rho=0\,$ of the
$\,n=4\,$ renormalization group. The natural question is whether
it is true that the effect of the Gauss-Bonnet term is to kill the
asymptotic freedom in $\,n=4-\ep\,$ dimensions. The answer to this
question is definitely not. The source of our failure to see the
standard fixed point is that we have used only the algebraic
equations $\,\be_\la=\be_\rho=0\,$ and due to the non-polynomial
form of the $\,\be_\la$-function (\ref{beta 2}) one can not see
the fixed point with $\,\rho=0\,$ in this way. So, in order to
complete our study we have to consider, especially, the
possibility of simultaneous $\,\la\to 0\,$ and $\,\rho\to 0\,$.
Using elementary transformations, one can check that the regimes
$\,\la \ll \rho\,$ and $\,\rho \ll \la\,$ lead to contradictions.
Therefore, we consider, additionally, the possibility of the
special solution $\,\rho = k\la$, where $\,k\,$ is a constant.
Under this assumption, the equations (\ref{beta 2}) are consistent
if \beq f_1\, k^3+(2g_1 -f_2)\, k^2 + (f_3 - 2g_2)\, k + 2g_3 =
0\,, \label{criterium 1} \eeq with an additional restriction \beq
b\,=\,\frac{2}{(4\pi k)^2} \,\Big(g_1 k^2 - g_2k + g_3\Big) > 0\,,
\label{criterium 2} \eeq dictated by the asymptotic freedom, in
the UV regime. The origin of this condition is the following.
After the relation $\,\rho=k\cdot \la\,$ is imposed, the equation
for $\,\la\,$ becomes \beq
\frac{d\la}{dt}\,=\,-\ep\,\la\,-\,b\,\la^2\,. \label{criterium 3}
\eeq The general solution of this equation has the form \beq
\la(t)\,=\,\frac{\ep}{b\,(e^{\ep t} \,-\, 1)}\,, \qquad\quad
\la(t_0)>0\,. \label{criterium 4} \eeq It is easy to see that the
asymptotic freedom in the UV limit $\,t\to +\infty\,$ holds for
$\,\ep > 0\,$ and (\ref{criterium 2}) satisfied. For $\,\ep < 0\,$
and condition (\ref{criterium 2}) satisfied the situation is more
complicated, because the UV stable fixed point is non-zero
$\la(t\to\infty) = -\ep/b > 0$. However, this fixed point tends to
zero when $\ep\to 0$, and we can consider that the theory is
asymptotically free in this sense. At the same time, independent
on the sign of $\,\ep$, the theory with $\,b<0\,$ does not
manifest the asymptotically free behavior in the UV limit.

In fact, there is no guarantee that the condition
(\ref{criterium 2}) is satisfied for every choice of
$\ep$ and all real roots of (\ref{criterium 1}).
The numerical computations show that for
$\,\ep=-0.01\,\, ,\, -0.1\,$ and $\,-1$, the
equation (\ref{criterium 1}) has only one real root
and that this root satisfies the condition
(\ref{criterium 2}).
On the other side, for $\ep=0.9\, ,\, 0.1$ and $0.01$,
the equation (\ref{criterium 1}) has three distinct real
roots, one of them violating condition (\ref{criterium 2})
for each case. Thus, there are solutions
of the equations $\,\rho = k\la\,$ and (\ref{criterium 1})
which do not satisfy Eq. (\ref{criterium 2}). Let us
remark that, in all cases we have examined, there are
also solutions with the UV stable fixed point $(0,0)$.
However, the asymptotic freedom depends on the choice
of the initial condition on the $\,\la-\rho\,$ plane.
In some cases, when Eq. (\ref{criterium 2}) does not
hold on the special solution of the equations
(\ref{beta 2}), the $(0,0)$-point is not stable in the UV.

Looking at the Figures 1-4, one can observe the
renormalization group trajectories (for the $\ep >0$
case) linking the IR-stable point $(\la_3,\rho_3)$ to
the UV-stable point $(0,0)$, or alternatively the
IR-stable point $(\la_4,\rho_4)$ to $(0,0)$. The
situation is similar for $\ep <0$,
but here the renormalization
group flow is inverted, linking the IR-stable point $(0,0)$
to the UV-stable point $(\la_4,\rho_4)$.

\section{\label{6}Conclusions and discussions}

We have calculated the one-loop effective action for the Weyl
gravity with the Gauss-Bonnet term. In the $\,n\to 4\,$ limit the
quantum effects of the Gauss-Bonnet term cancel. This cancellation
may be seen as a negative answer to the problem raised in
\cite{capkim}. This result is valid, at least, in the framework of
the conformal quantum gravity. Another remarkable fact is that, in
agreement with \cite{amm}, there is no infinite
$\,\int\sqrt{-g}R^2\,$ counterterm. Other sectors of the divergent
part of the effective action are in a perfect agreement with both
earlier calculations \cite{frts82,amm}.

Despite the one-loop divergences are conformal invariant,
this symmetry may be broken at the one-loop level in the
finite part of the effective action. The divergences of
the $\,\int\sqrt{-g}C^2\,$ and $\,\int\sqrt{-g}E$-type
produce the anomalous violation of the Noether identity
(\ref{Noether}), and as a result the finite part of the
one-loop effective action contains usual
non-local \footnote{The non-local
form of the anomaly-induced gravitational action has been
first indicated in \cite{ddi}.}
anomaly-induced terms  \cite{rei}. There may be
also local $\,\int\sqrt{-g}R^2$-type contribution
which deserves a special discussion. It is easy to see
that there are two different possible sources of this term
in the Weyl quantum gravity:
\vskip 1mm

i) If the calculation is performed in a dimensional
regularization, the $\,\de^{(0)}_3\,$ and $\,\de^{(1)}_3\,$
terms in
(\ref{coefficients}) are proportional to $\,(n-4)\,$
and therefore they produce the finite $\,\int\sqrt{-g}R^2$
term directly from $\,A_2^t$. It is remarkable
that this contribution depends on the coefficient $\,\eta\,$
of the Gauss-Bonnet term. According to \cite{anomayo}, this
contribution is a subject of strong ambiguity typical for
the dimensional regularization. In general, the dimensional
regularization is unable to predict any definite value for
the coefficient of the finite $\,\int\sqrt{-g}R^2$-term.
\vskip 1mm

ii) The infinite $\,\int\sqrt{-g}\square R\,$-type
counterterm, which we did not calculate here, may
produce contribution to the conformal anomaly and
eventually to the finite $\,\int\sqrt{-g}R^2$ term.
However, this contribution is plagued by a double
ambiguities. First of all, the
$\,\int\sqrt{-g}\square R\,$-type counterterm
itself is gauge-fixing dependent \cite{shja}.
As it was already explained above,
this is the reason why we did not calculate this
counterterm. The second source of
ambiguity is the derivation of anomaly and of the
anomaly-induced effective action. In relation to the
$\,\int\sqrt{-g}R^2\,$ term these procedures may be
ambiguous. The detailed discussions of this issue have
been given recently in \cite{anomayo}, where the
ambiguity has been confirmed not only for the traditional
version of the dimensional regularization (where it
is completely out of control) but also in a more
physical covariant Pauli-Villars regularization
with non-minimal scalar massive regulators. It is
worth noticing that the status of this last ambiguity
in the Weyl quantum gravity is quite different from
the one in the semiclassical approach. In the last
case the ambiguity is always reduced to the freedom
of adding the $\,\int\sqrt{-g}R^2\,$ term to the
classical action of vacuum, while in the former
case this operation would increase the number of physical
degrees of freedom (see, e.g. \cite{book} and references
therein) and hence can not be seen as the legal
operation for the theory (\ref{Weyl}).

In any case the local conformal invariance in Weyl gravity
is
violated at the one-loop level by quantum corrections. Hence,
despite the general higher derivative quantum gravity is indeed
renormalizable \cite{stelle,tyutin}, the particular conformal
version is multiplicatively non-renormalizable at higher loops.
Our results show, however, that the conformal quantum gravity can
be regarded as a good approximation. The corresponding procedure
means that one can start from the theory with a very small
coefficient of the $\,\int\sqrt{-g}R^2\,$ term. Due to the
one-loop renormalizability of the conformal theory this
coefficient will remain very small at the quantum level. If we
consider the conformal quantum gravity in this framework, the
problem of ambiguity of the anomalous $\,\int\sqrt{-g}R^2\,$ term
is irrelevant and we can regard this theory as a useful particular
example of the higher derivative quantum gravity models.

One of the outputs of our investigation are new fixed
points of the renormalization group flows which appear due
to quantum effects of the topological Gauss-Bonnet term
in $\,4-\ep\,$  dimension \footnote{
The non-trivial fixed points for the Einstein gravity,
including the case with the higher derivative truncation
were reported earlier in \cite{reuter}. It would be
interesting to discuss the possible relation between the
two kind of non-trivial fixed points. This relation may
take place for the general higher derivative quantum
gravity which has a proper Einstein-Hilbert low-energy
limit.}. One can expect even greater number of non-trivial
fixed points for a general higher derivative quantum
gravity, with the Einstein-Hilbert, cosmological and
$\,\int R^2$-terms included.

\vskip 4mm

\noindent
{\large\bf Acknowledgments.}
One of the authors (I.Sh.) is grateful to I.L. Buchbinder
and I.V. Tyutin for numerous discussions of the Weyl quantum
gravity in the period between 1981 and 1993. The work of
the authors has been supported by the research grant from
FAPEMIG and by the fellowships from FAPEMIG (G.B.P.)
and CNPq (I.Sh.).

\appendix
\section{\label{A} Bilinear
expansions quadratic in curvature terms}

In this Appendix we collect the cumbersome expressions
with the bilinear expansions of the relevant terms of
the second order in curvature. Furthermore we
present the transformation of these terms to the form
which is useful for the derivation of the effective
action.
The initial set of the bilinear expansions has
the following form:
\begin{widetext}
\beq
\Big(\sqrt{-g}\, R^2_{\mu\nu\al\be}\Big)^{(2)}
& = & \sqrt{-g}h^{\mu\nu}\,\Big\{\,
\de_{\mu\nu\, ,\,\al\be}\Box ^2
- g_{\mu\al}\na_\be\Box\na_\nu
+ \na_\al\na_\be\na_\mu\na_\nu -
g_{\mu\al}\na_\rh\na_\be\na_\nu\na^\rh +
4R^\rh\mbox{}_{\al\nu\be}\na_\mu\na_\rh + \nonumber
\\
& + & \de_{\mu\nu\, ,\,\al\be}R_{\rh\la}\na^\rh\na^\la  +
R_{\mu\al\nu\be}\Box - 2R^\rh\mbox{}_{\mu\al\nu}\na_\be\na_\rh
- 2g_{\mu\nu}R_{\rh\al\la\be}\na^\la\na^\rh -
4g_{\nu\be}R_{\al\rh\la\mu}\na^\la\na^\rh + \nonumber
\\
& + &
\frac{7}{2}g_{\nu\al}R_{\mu\rh\la\ta}R_{\be}\mbox{}^{\rh\la\ta}
+ g_{\mu\nu}R_{\rh\la\ta\al}R^{\rh\la\ta}\mbox{}_\be
- \frac{1}{4}\,
\Big(\de_{\mu\nu\, ,\,\al\be}-\frac12\,g_{\mu\nu}g_{\al\be}\Big)
R_{\rh\la\ta\th}^2
- \frac{1}{2}R_{\mu\al\rh\la}R_{\nu\be}\mbox{}^{\rh\la}\,
 \Big\}\,h^{\al\be}\, ; \nonumber
\eeq
\beq \Big(\sqrt{-g}\, R^2_{\mu\nu}\Big)^{(2)} & = & \sqrt{-g}\,
h^{\mu\nu}\,\Big\{ \frac{1}{2}\na_\al\na_\mu\na_\be\na_\nu
+\frac{1}{4}\de_{\mu\nu\, ,\,\al\be}\Box ^2 +
\frac{1}{2}g_{\nu\al}\na_\rh\na_\mu\na_\be\na^\rh -
\frac{1}{2}g_{\al\be}\na_\rh\na_\mu\na_\nu\na^\rh -
\nonumber \\
& - & \frac{1}{2}g_{\mu\nu}\na_\al\na_\rh\na_\be\na^\rh
+ \frac{1}{4} g_{\mu\nu}g_{\al\be}\na_\rh\Box\na^\rh -
\frac{1}{2}g_{\al\nu}\na_\be\na_\mu\Box +\frac{1}{2}
g_{\mu\nu}\na_\al\na_\be\Box
- \frac{1}{2}g_{\nu\al}\Box\na_\be\na_\mu -
\nonumber \\
& - & 2g_{\nu\al}R^\rh\mbox{}_\be\na_\mu\na_\rh +
\frac{1}{2}\de_{\mu\nu\,,
\,\al\be}R^{\rh\la}\na_\rh\na_\la +
g_{\al\be}R^\rh\mbox{}_\nu\na_\rh\na_\mu
- R_{\mu\be}\na_\al\na_\nu + \nonumber
\\
& + & g_{\nu\al}R_{\mu\be}\Box+
2g_{\nu\al}R^\rh\mbox{}_\mu\na_\rh\na_\be +
g_{\mu\nu}R^\rh\mbox{}_\be\na_\al\na_\rh -
\frac{1}{2}g_{\mu\nu}g_{\al\be}R^{\rh\la}\na_\rh\na_\la -
2g_{\mu\be}R^\rh\mbox{}_\nu\na_\al\na_\rh +
\nonumber \\
& + & \frac{1}{8}(g_{\mu\nu}g_{\al\be} - 2\de_{\mu\nu
\,,\,\al\be})R_{\rh\la}^2 + R_{\mu\al}R_{\nu\be} +
2g_{\nu\al}R_{\mu\rh}R^\rh\mbox{}_\be -
g_{\al\be}R_{\mu\rh}R^\rh\mbox{}_\nu \Big\}\, h^{\al\be}\, ;
\eeq
and
\beq
\Big(\sqrt{-g}\, R^2\Big)^{(2)} & = & \sqrt{-g}\,
h^{\mu\nu}\, \Big\{\na_\mu\na_\nu\na_\al\na_\be -
g_{\al\be}\na_\mu\na_\nu\Box - g_{\mu\nu}\Box\na_\al\na_\be +
g_{\mu\nu}g_{\al\be}\Box ^2 - g_{\nu\al}R\na_\be\na_\mu -
\nonumber \\
& - & 2R_{\mu\nu}\na_\al\na_\be + g_{\mu\nu}R\na_\al\na_\be +
2g_{\al\be}R_{\mu\nu}\Box
+ \frac{1}{2}(\de_{\mu\nu\, ,\,\al\be} -  g_{\mu\nu}g_{\al\be})R\Box
+ \nonumber \\
& + &  2Rg_{\nu\be}R_{\mu\al} -
g_{\mu\nu}RR_{\al\be} + \frac{1}{8}(g_{\mu\nu}g_{\al\be}
- 2\de_{\mu\nu\,,\,\al\be})R^2 + R_{\mu\nu}R_{\al\be}\Big\}
\, h^{\al\be}\,.
\eeq
\end{widetext}

It proves necessary to establish some commutation relations
between covariant derivatives. In the expressions below we have
omitted those terms which may contribute only to the total
derivatives in the effective action. Also, for the sake of brevity
we broke the symmetries in the pairs of indices $\,\,(\al\be)\,\,$
and $\,\,(\mu\nu)$. These symmetries has to be restored for
practical calculations. We can write \beq
g_{\nu\be}\na^\la\na_\mu\na_\al\na_\la\, h^{\al\be} & = &
(g_{\nu\be}R_{\rh\mu}\na^\rh\na_\al +
R_{\nu\be\la\mu}\na^\la\na_\al + \nonumber \\
& + &
g_{\nu\be}\na_\mu\Box\na_\al +
g_{\nu\be}R^\la\mbox{}_\al\na_\la\na_\mu +
\nonumber \\
& + & R_{\nu\be\al\la}\na^\la\na_\mu )\, h^{\al\be}\, , \nonumber
\eeq
\beq
g_{\nu\be}\Box\na_\al\na_\mu\, h^{\al\be} & = &
(g_{\nu\be}R_{\mu\al}\Box +R_{\nu\be\al\mu}\Box +
\nonumber \\
& + & g_{\nu\be}\na_\mu\Box\na_\al
+ g_{\nu\be}R_{\rh\mu}\na^\rh\na_\al +
\nonumber \\
& + & 2R_{\nu\be\la\mu}\na^\la\na_\al )\, h^{\al\be}\, , \nonumber
\eeq
\beq
g_{\nu\be}\na^\la\na_\al\na_\mu\na_\la\, h^{\al\be} & =
& (g_{\nu\be}R_{\rh\mu}\na^\rh\na_\al +
R_{\nu\be\la\mu}\na^\la\na_\al +
\nonumber \\
& + & g_{\nu\be}\na_\mu\Box\na_\al +
g_{\nu\be}R^\la\mbox{}_\al\na_\la\na_\mu +
\nonumber \\
& + & R_{\nu\be\al\la}\na^\la\na_\mu +
g_{\nu\be}R_{\mu\al}\Box + \nonumber
\\
& + & R_{\nu\be\al\mu}\Box +
\frac{1}{2}g_{\nu\be}R_{\mu\ta\rh\la}R_\al\mbox{}^{\ta\rh\la}
+ \nonumber
\\
& + & \frac{1}{2}R_{\al\mu}\mbox{}^{\rh\la}R_{\nu\be\rh\la}) \,
h^{\al\be}\, , \nonumber
\eeq
\beq
g_{\nu\be}\na_\al\Box\na_\mu\,
h^{\al\be} & = &
(2g_{\nu\be}R^\rh\mbox{}_{\al\la\mu}\na_\rh\na^\la + \nonumber
\\
& + & 2R_{\nu\be\la\mu}\na_\al\na^\la +
R_{\mu\al\be\nu}\Box + \nonumber
\\
& + & g_{\nu\be}(2R_{\rh\al}\na_\mu\na^\rh
+R_{\al\mu}\Box + \nonumber
\\
& + & \na_\mu\Box\na_\al + 2R_{\be\nu\la\al}\na_\mu\na^\la )\,
h^{\al\be}\, , \nonumber
\eeq
\beq
g_{\al\be}\na_\rh\na_\la\na^\rh\na^\la\, h^{\al\be} & = &
(R_{\rh\la}\na^\rh\na^\la +\Box^2)\, h , \nonumber
\eeq
\beq
\na_\al\na_\mu\na_\be\na_\nu\, h^{\al\be} & = &
(R^\rh\mbox{}_{\al\be\nu}\na_\rh\na_\mu + R_{\nu\be}\na_\al\na_\mu
+ \nonumber
\\
& + & 2R_{\mu\al}\na_\nu\na_\be -
R^\rh\mbox{}_{\nu\al\mu}\na_\rh\na_\be +
\nonumber \\
& + & \na_\mu\na_\nu\na_\al\na_\be )\, h^{\al\be}\, , \nonumber
\eeq
\beq
\na_\al\na_\be\na_\mu\na_\nu\, h^{\al\be} & = &
\Big[\,\na_\mu\na_\nu\na_\al\na_\be + 4R_{\mu\al}\na_\nu\na_\be -
\nonumber \\
& - & R^\rh\mbox{}_{\nu\al\mu}(\na_\rh\na_\be
+ \na_\be\na_\rh ) +
\nonumber \\
& + &  2R^\rh\mbox{}_{\al\be\mu}\na_\rh\na_\nu \,\Big]\,
h^{\al\be}\, , \nonumber
\eeq
\beq
\na^\la\na_\mu\na_\la\na_\nu\,
h & = & (R_{\mu\rh}\na^\rh\na_\nu - R_{\rh\nu\la\mu}\na^\la\na^\rh
+ \nonumber \\
& + & \na_\mu\Box\na_\nu )\, h\, , \nonumber
\\
\Box\na_\mu\na_\nu\, h & = &
(2R^\rh\mbox{}_{\mu\nu}\mbox{}^\la\na_\rh\na_\la +
R_{\rh\mu}\na^\rh\na_\nu +
\nonumber \\
& + & \na_\mu\Box\na_\nu )\, h\,, \nonumber
\eeq
\beq
\na_\al\na^\la\na_\be\na_\la\, h^{\al\be} & = &
(R_{\rh\be}\na_\al\na^\rh +
R^\rh\mbox{}_{\al\be}\mbox{}^\la\na_\rh\na_\la
+ \nonumber \\
& + & \na_\al\Box\na_\be )\, h^{\al\be}\, .
\label{commutator 2}
\eeq
Using these relations, we can rewrite the bilinear
expansions in a more useful form
\begin{widetext}
\beq
\Big(\sqrt{-g}\, R^2_{\mu\nu\al\be}\Big)^{(2)}
& = & \sqrt{-g}\, h^{\mu\nu}\,\left\{\,
\de_{\mu\nu\, ,\,\al\be}\Box^2
+ g_{\nu\be}R_\al\mbox{}^{\rh\la}\mbox{}_\mu (2\na_\rh\na_\la
-4\na_\la\na_\rh ) \,-
R_{\nu\be\la\mu} (\na^\la\na_\al -
2\na_\al\na^\la )+ \right. \nonumber
\\
& + & \left. 3R_{\mu\al\nu\be}\Box
+ 5 R_{\rh\al\nu\be}\na_\mu\na^\rh \,-
g_{\nu\be}R_{\rh\al}(\na_\mu\na^\rh + \na^\rh\na_\mu )
\,-\,g_{\nu\be}\,R_{\rh\mu}(\na_\al\na^\rh + \na^\rh\na_\al )-
\right.\nonumber \\
& - & \left. 2g_{\nu\be}\na_\mu\Box\na_\al -
2g_{\nu\be}R_{\mu\al}\Box
+ \na_\mu\na_\nu\na_\al\na_\be
\,+\,3g_{\nu\be}R_{\mu\ta\rh\la}R_\al\mbox{}^{\ta\rh\la}\,+
\de_{\mu\nu\, ,\,\al\be}R_{\rh\la}\na^\rh\na^\la - \right.
\nonumber \\
& - & \left. g_{\mu\nu} (2R_{\rh\al\la\be}\na^\la\na^\rh \,+
R_{\rh\la\ta\al}R^{\rh\la\ta}\mbox{}_\be ) \,+
\frac{1}{8} R_{\rh\si\la\tau}^2(g_{\mu\nu}g_{\al\be} -
2\de_{\mu\nu\, ,\,\al\be})+ \right. \nonumber
\\
& + & \left. 2R_{\mu\nu} (\na_\nu\na_\be +\na_\be\na_\nu ) \right\}\,
h^{\al\be}\, ;
\label{Riemann}
\eeq
\beq
\Big(\sqrt{-g}\, R^2_{\mu\nu}\Big)^{(2)}
& = & \sqrt{-g}\, h^{\mu\nu}\,\Big\{\,
\frac{1}{2}\na_\mu\na_\nu\na_\al\na_\be
- \frac{1}{2}g_{\mu\nu}\na_\al\Box\na_\be -
\frac{1}{2}g_{\nu\be}\na_\mu\Box\na_\al +
\frac{1}{4}(\de_{\mu\nu\, ,\,\al\be}
+ g_{\mu\nu}g_{\al\be})\Box^2 +
\nonumber \\
& + & \frac{1}{2}R_{\nu\be}\na_\al\na_\mu
-\frac{1}{2}g_{\nu\be}R_{\rh\mu}(\na^\rh\na_\al
+ 3\na_\al\na^\rh ) +
\frac{3}{2}g_{\al\be}R_{\rh\mu}\na^\rh\na_\nu
- R_{\mu\al\be\nu}\Box+
\nonumber \\
& + & \frac{1}{4}(2\de_{\mu\nu\, ,\,\al\be}
- g_{\mu\nu}g_{\al\be})R^{\rh\la}\na_\rh\na_\la
+
R_{\la\mu\nu\al}R_\be\mbox{}^\la
+ R_{\mu\rh\la\nu}R_\al\mbox{}^{\rh\la}\mbox{}_\be
- \frac{1}{8}(2\de_{\mu\nu\, ,\,\al\be}
- g_{\mu\nu}g_{\al\be})R_{\rh\la}^2 +
\nonumber \\
& + &  R_{\mu\al}R_{\nu\be}
+ 2g_{\nu\be}R_{\mu\rh}R^\rh\mbox{}_\al
- g_{\al\be}R_{\mu\rh}R^\rh\mbox{}_\nu \Big\}\, h^{\al\be}\, ;
\label{Ricci}
\eeq
\beq
\Big(\sqrt{-g}\, R^2\Big)^{(2)}
& = & \sqrt{-g}\, h^{\mu\nu}\,\Big\{\,
\na_\mu\na_\nu\na_\al\na_\be- 2g_{\al\be}\na_\mu\Box\na_\nu
+ g_{\al\be}R_{\la\nu}\na_\mu\na^\la \,+
g_{\mu\nu}R_{\la\be}\na^\la\na_\al
+g_{\mu\nu}g_{\al\be}\Box^2 -
\nonumber \\
& - & g_{\nu\be}R\na_\al\na_\mu\, -
2R_{\mu\nu}\na_\al\na_\be \, +
2g_{\al\be}R_{\mu\nu}\Box
+ g_{\mu\nu}R\na_\al\na_\be
 + \frac{1}{2}(\de_{\mu\nu\,,\,\al\be}
- g_{\mu\nu}g_{\al\be})R\Box -
\nonumber \\
& - & g_{\mu\nu}RR_{\al\be}\, +
\frac{1}{8}(g_{\mu\nu}g_{\al\be}
-2\de_{\mu\nu\, ,\,\al\be})R^2 +2Rg_{\nu\be}R_{\mu\al}
+ R_{\mu\nu}R_{\al\be}
\Big\}\, h^{\al\be}\, .
\label{scalar}
\eeq
\\
\end{widetext}
\section{\label{B} Particular
results of the calculations in the background field
method}

In this Appendix one can find the results for the particular
elements of the expression (\ref{gr-formula}). One can easily find
the contribution of the commutator (\ref{commute})
\begin{eqnarray}
\frac{1}{6}\,{\rm tr} \,\hat{\cal R}_{\mu\nu}\hat{\cal R}^{\mu\nu}
= -\frac{n+2}{6}\,R^2_{\mu\nu\al\be}\,.
\end{eqnarray}
After a
tedious algebra, we arrive at the following result:
\begin{widetext}
\beq
{\rm tr}\, \hat{U} = aR_{\mu\nu\al\be}^2 + bR_{\mu\nu}^2 +
cR^2\,, \eeq where \beq \left(\begin{array}{lll} a \\ b \\ c \cr
\end{array}\right)\,\,=\,\, \frac{1}{2n(y+4x)}\,
\left(\begin{array}{lll} 5xn^2+26xn-24x-xn^3+6yn
\\ 5yn^2+10yn-24y-yn^3+24xn+8zn
\\ 2y n+4xn+6zn-24z+5zn^2-zn^3
\cr \end{array}\right)\,_.
\eeq

Other relevant traces are the following:
\beq
{\rm tr}\, \left(R\hat{V}^\rh\mbox{}_\rh\right)=
(n+2)\left[\,(n-1)(a_1+na_2)- a_3\,\right] R^2\, ,
\nonumber \\
\eeq
where
\beq
\left(\begin{array}{lll}
a_1 \\ a_2 \\ a_3 \cr \end{array}\right)\,\,=\,\,
\frac{2}{n\,(y+4x)}\,
\left(\begin{array}{lll}
nb_1+b_4+b_8-b_9
\\
nb_2+b_5+b_6
\\
nb_3-b_7 \cr \end{array}\right)\,; \nonumber \eeq also \beq {\rm
tr}\, \left(R_{\rh\si}\hat{V}^{\rh\si}\right) & = & a_4\,
R_{\mu\nu}^2 \,+ \,a_5\,R^2\,,
\eeq
where
\beq
\left[\begin{array}{ll} a_4 \\ a_5  \cr \end{array}\right] =
\frac{2}{n(y+4x)} \left[\begin{array}{ll}
(n-2)b_4+(n^2+n-2)(nb_6+b_8-b_9)+ (n+2)b_7 +2nb_{10}
\\
(n^2+n-2)(b_1+nb_2+b_5)-
(n+2)b_3+nb_4-2b_{10}
\cr \end{array}\right];
\nonumber
\eeq
\beq
{\rm tr}\,
\left( \hat{V}^\rh\mbox{}_\rh\right)^2
& = &
12\,{\frac {\left(3\,nx+ny+4\,x\right)^{2}}
{\left(y+4\,x\right)^{2}}}\, R_{\mu\nu\al\be}^2 +
\frac{16}{\left (y+4\,x\right )^{2}n}\left\{\,
{n}^{4}{x}^{2}+8\,{n}^{3}{x}^{2}+6\,{n}^{2}{x}^{2}+
4\,{n}^{3}xy +6\,{n}^{2}xy + {y}^{2}{n}^{2}
+\right.
\nonumber
\\
& + & \left.
10\,{y}^{2}n- 8\,{y}^{2}-16\,n{x}^{2}
-32\,{x}^{2}+32\,yx\,\right\}\, R_{\mu\nu}^2 +
\frac{2}{\left (y+4\,x\right )^{2}{n}^{2}}\left\{\,
-24\,{n}^{3}xy - 48\,{n}^{2}xy+32\,nxy- \right.
\nonumber
\\
& - & \left.
8\,{z}^{2}{n}^{2}+ 12\,{z}^{2}{n}^{3}+
{n}^{6}{z}^{2}-3\,{n}^{5}{z}^{2}+
{n}^{4}{y}^{2}-11\,{n}^{3}{y}^{2} -128\,yx -
48\,{n}^{2}{x}^{2}+
32\,{y}^{2}-4\,{n}^{4}{x}^{2}-36\,{n}^{3}{x}^{2}-
\right. \nonumber
\\
& - & \left. 2\,{y}^{2}{n}^{2}-16\,{y}^{2}n - 14\,{n}^{4}zy +
128\,{x}^{2} -2\,{z}^{2}{n}^{4} -4\,{n}^{5}zx-8\,{n}^{4}zx-
32\,zny+ 56\,z{n}^{2}y + \right.
\nonumber
\\
& + & \left. 2\,{n}^{5}zy + 64\,znx
+ 32\,z{n}^{2}x-4\,{n}^{4}xy \,\right\}\,R^2\,;
\eeq
and finally
\beq
{\rm tr}\, \left(\hat{V}_{\rh\si}\hat{V}^{\rh\si}\right)
& = &
c_1\, R_{\mu\nu\al\be}^2 +c_2\, R_{\mu\nu}^2+c_3\, R^2\, ,
\eeq
where the constants $c_1$, $c_2$ and $c_3$ are given by
\beq
c_1 & = &
{\frac {12\,(24\,n{x}^{2}+8\,nxy+12\,{n}^{2}{x}^{2}-32\,{x}^{2}
+6\,{n}^{2}xy+{y}^{2}{n}^{2})}{\left (y+4\,x\right )^{2}n}}\, ,
\eeq
\beq
c_2 & = &
\frac{2}{\left (y+4\,x\right )^{2}{n}^{2}}\left\{\,
24\,{n}^{4}{x}^{2}+96\,{n}^{3}{x}^{2}+72\,{n}^{2}{x}^{2}-
192\,n{x}^{2}+48\,{y}^{2}n-
10\,{y}^{2}{n}^{2}-128\,yx-192\,nxy+ \right.
\nonumber
\\
& + & \left. 40\,{n}^{2}xy-32\,z{n}^{3}x+128\,znx+
32\,z{n}^{2}x-64\,zny
+ 16\,z{n}^{3}y+ 16\,z{n}^{2}y
+ 52\,{n}^{3}xy+12\,{n}^{4}xy+\right.
\nonumber
\\
& + & \left.
128\,{x}^{2} + 3\,{n}^{4}{y}^{2} + {n}^{3}{y}^{2}
+8\,{z}^{2}{n}^{4}+8\,{z}^{2}{n}^{3}+32\,{y}^{2}\,\right\}
\eeq
and
\beq
c_3 & = &
\frac{4}{\left (y+4\,x\right )^{2}{n}^{2}}\left\{\,
-4\,{n}^{3}xy+2\,{n}^{2}xy-4\,nxy+8\,z{n}^{3}x-z{n}^{3}y -
4z^2n^2- 9\,{z}^{2}{n}^{3}+{n}^{5}{z}^{2}+\right.
\nonumber
\\
& + & \left. {n}^{3}{y}^{2}+32\,{
x}^{2}+16\,yx-16\,{n}^{2}{x}^{2}-
16\,{y}^{2}-8\,{n}^{3}{x}^{2}-8\,{y}^{2}n-4\,n{x}^{2}+
3\,{n}^{4}zy-4\,zny-48\,znx+ \right.
\nonumber
\\
& + & \left. 24\,z{n}^{2}x-24\,z{n}^{2}y
+ 4\,n{z}^{2}+48\,zy-96\,zx\,\right\}\, .
\eeq
\end{widetext}




\begin{thebibliography}{99}

\bibitem{birdav} N.D. Birrell and P.C.W. Davies,
{\sl Quantum fields in curved space}
(Cambridge University Press, 1982).


\bibitem{book}
I.L. Buchbinder, S.D. Odintsov and I.L. Shapiro, {\sl
Effective Action in Quantum Gravity} (IOPP, Bristol, 1992).

\bibitem{duff94}
M.J. Duff, Class. Quant. Grav. {\bf 11} (1994) 1387.

\bibitem{rei} R.J. Riegert, Phys. Lett. {\bf B 134} (1984) 56;

E.S. Fradkin and A.A. Tseytlin, Phys.Lett. {\bf 134B} (1984) 187.

\bibitem{cofe} G. de Berredo-Peixoto and I.L. Shapiro,
Phys. Lett. {\bf B 514} (2001) 377.

\bibitem{frts-sugra} E.S. Fradkin and A.A. Tseytlin,
Phys. Repts. {\bf 119} (1985) 233.

\bibitem{hathrell} S.J. Hathrell, Ann. Phys. (NY) {\bf 139}
(1982) 136, {\bf 1142} (1982) 34.

\bibitem{anomayo} M. Asorey, E.V. Gorbar, I.L. Shapiro,
Class. Quant. Grav. {\bf 21} (2003) 163. 

\bibitem{wald} R.M. Wald, General Relativity. (University of
Chicago Press, 1984).

\bibitem{frts82} E.S. Fradkin and  A.A. Tseytlin,
Nucl. Phys. {\bf B201} (1982) 469.

\bibitem{julve}
J. Julve and M. Tonin, Nuovo Cim. {\bf 46 B} (1978) 137.

\bibitem{avba}
I.G. Avramidi, Sov. J. Nucl.Phys. {\bf 44} (1986) 255;
I.G. Avramidi, A.O. Barvinsky, Phys.Lett. {\bf 159B} (1985) 269;
I.G. Avramidi, Ph.D. thesis. [hep-th/9510140].

\bibitem{truffin} F. Englert, C. Truffin and R. Gastmans,
Nucl.Phys. {\bf 117B} (1976) 407.

\bibitem{frvi}  E.S. Fradkin and G.A. Vilkovisky,
Phys. Lett. {\bf 73B} (1978) 209.

\bibitem{bush86} I.L. Buchbinder and I.L. Shapiro,
Yad. Fiz. (Sov. J. Nucl. Phys.) {\bf 44} (1986) 1033.

\bibitem{shja} I.L. Shapiro and A.G. Jacksenaev,
Phys. Lett. {\bf 324B} (1994) 286.

\bibitem{amm} I. Antoniadis, P.O. Mazur and E. Mottola,
Nucl. Phys. {\bf B388} (1992) 627.

\bibitem{capkim} D.M. Capper and D. Kimber,
J. Phys. {\bf A13} (1980) 3671.

\bibitem{stelle}
K.S. Stelle,  Phys. Rev. {\bf D16} (1977) 953.

\bibitem{Christen} N.H. Barth and S.M. Christensen,
Phys. Rev. {\bf D28} (1983) 1876.

\bibitem{MAPLE} A. Heck,
{\sl Introduction to Maple}, 
(Springer-Verlag, 2003).

\bibitem{salam} A. Salam and J. Strathdee, Phys. Rev. {\bf D 18}
(1978) 4480.

\bibitem{shap}
I.L. Shapiro, Class. Quant. Grav. {\bf 6} (1989) 1197.

\bibitem{BKSVW} I.L. Buchbinder, O.K. Kalashnikov,  I.L. Shapiro,
V.B. Vologodsky and Yu.Yu. Wolfengaut,
Phys. Lett. {\bf 216B} (1989) 127;
Fortschr. Phys. {\bf 37} (1989) 207.

\bibitem{GKT} R. Gastmans, R. Kallosh and C. Truffin,
Nucl. Phys. {\bf B133} (1978) 417;

S.M. Christensen and M.J. Duff, Phys. Lett.
{\bf B79} (1978) 213.

\bibitem{Weinberg 2-e} S. Weinberg, in: {\sl General Relativity.}
ed: S.W. Hawking and W. Israel, (Cambridge University Press, 1979).

\bibitem{KN} H. Kawai and M. Ninomiya, Nucl. Phys.
{\bf B336} (1990) 115;


H. Kawai, Y. Kitazawa and M. Ninomiya, Nucl.Phys.
{\bf B404} (1993) 684; {\bf B393} (1993) 280;



S. Kojima, N. Sakai and Y. Tanii,
{\sl Nucl.Phys.} {\bf 426B} 223 (1994).

\bibitem{ddi} S. Deser, M.J. Duff and C. Isham,
Nucl. Phys. {\bf 111B} (1976) 45.

\bibitem{tyutin} B.L. Voronov and I.V. Tyutin,
Yad.Fiz. (Sov. Journ. Nucl. Phys.) {\bf 39} (1984) 998.

\bibitem{reuter}
O. Lauscher and M. Reuter,
Phys.Rev. {\bf D66} (2002) 025026; {\bf D65} (2002) 025013.




\end{thebibliography}
\end{document}